\documentclass[a4paper,12pt]{article}
\usepackage{jheppub}
\usepackage{placeins,float}
\usepackage{graphicx,dcolumn, bm,epsfig, cancel}
\usepackage[utf8]{inputenc}
\usepackage{amsmath}
\numberwithin{equation}{section}
\usepackage{amssymb, times}
\usepackage{fontspec}
\usepackage{color}
\usepackage{makecell}
\usepackage{booktabs}
\usepackage{tabularx}
\usepackage[title,titletoc]{appendix}
\hypersetup{hypertexnames=false}
\renewcommand{\figurename}{Fig.}
\makeatletter
\gdef\@fpheader{}
\makeatother

\newcommand{\pkg}[1]{\texttt{#1}}
\newcommand{\trianglesim}{\texttt{Triangle\discretionary{-}{}{-}Simulator}}

\title{Full-Covariance Bayesian Inference of Stochastic Gravitational-Wave Backgrounds with Time-Domain Simulations for Taiji-like Missions}

\author[a,b,c]{Qingyuan Liang}
\emailAdd{qingyuan211045@gmail.com}

\author[a,b]{Ju Chen}
\emailAdd{chenju@ucas.ac.cn}

\author[d]{Minghui Du}
\emailAdd{duminghui@imech.ac.cn}

\author[a,b]{Huai-Ke Guo}
\emailAdd{guohuaike@ucas.ac.cn} 

\affiliation[a]{International Centre for Theoretical Physics Asia-Pacific (ICTP-AP),\\ University of Chinese Academy of Sciences (UCAS), Beijing, China.}
\affiliation[b]{Taiji Laboratory for Gravitational Wave Universe (Beijing/Hangzhou), University of Chinese Academy of Sciences (UCAS), Beijing, China.}
\affiliation[c]{Department of Statistics, University of Auckland, 38 Princes Street, Auckland, New Zealand}
\affiliation[d]{Center for Gravitational Wave Experiment, National Microgravity Laboratory,\\ Institute of Mechanics, Chinese Academy of Sciences, Beijing, China.}

\abstract{
For Taiji-like missions, we implement a Bayesian spectral inference framework that combines second-generation time-domain (TD) simulations of time-delay interferometry (TDI) with a frequency-domain (FD) spectral likelihood for stochastic gravitational-wave background (SGWB) analyses. The \(X,Y,Z\) Michelson streams generated with \trianglesim{} are divided into finite segments, Fourier transformed, and modeled with a segment-dependent complex \(3\times3\) covariance matrix. For each segment we evaluate the orbit-dependent response functions and noise transfer functions, allowing unequal-arm and time-evolving effects to enter through the full \(XYZ\) covariance. Controlled simulations performed with \trianglesim{} show that the calculated functions reproduce the realization-averaged spectra at the few-percent level over the retained frequency band away from TDI nulls. We then compare parameter-estimation results for static equal-arm FD, equal-arm TD, and unequal-arm TD configurations, using in each case a full \(XYZ\)-covariance likelihood matched to the corresponding detector configuration. All three yield consistent uncertainty trends and Bayesian-evidence diagnostics for astrophysical-background recovery after marginalizing over instrumental noise and an effective Galactic double-white-dwarf foreground. Finally, in a ten-parameter model containing instrumental noise, an effective Galactic double-white-dwarf foreground, a stochastic astrophysical background, and a sound-wave spectrum from a cosmological first-order phase transition, we recover its peak amplitude and frequency and find Bayesian evidence favoring its inclusion in all three matched configurations.
}

\begin{document}

\makeatletter
\renewcommand\thefootnote{\arabic{footnote}} 
\makeatother

\maketitle

\section{Introduction}

Stochastic gravitational-wave (GW) backgrounds (SGWBs) are key science targets for millihertz space-based interferometers because they can probe both unresolved astrophysical source populations and physical processes in the early Universe \cite{Allen:1997ad,Romano:2016dpx,LISA:2017pwj,Caprini:2019egz}. As a representative cosmological benchmark, we consider the sound-wave (SW) contribution generated during a first-order phase transition (FOPT). This component is sourced by bulk plasma motion driven by expanding bubbles and can peak in the millihertz band for transitions near the electroweak scale \cite{Caprini:2015zlo,Hindmarsh:2015qta,Hindmarsh:2017gnf,Weir:2017wfa,Mazumdar:2018dfl,Bian:2021ini,Athron:2023xlk}. In realistic data, however, a cosmological component would not be inferred in isolation. Instrumental noise, the unresolved Galactic double-white-dwarf (DWD) foreground and stochastic compact-binary backgrounds can occupy the same frequency range and partially mimic one another in spectral shape \cite{Nelemans:2001hp,Robson:2018ifk,Regimbau:2011rp,Rosado:2011kv}. For Taiji-like missions, the relevant SGWB inference problem is therefore one of multi-component spectral separation rather than the recovery of a single idealized stochastic spectrum.

Taiji is a heliocentric triangular space-based interferometer concept with an architecture closely related to that of the Laser Interferometer Space Antenna (LISA) \cite{LISA:2017pwj,Hu:2017mde,Ruan:2018tsw,Wu:2018clg}. In the numerical setup considered here, the Taiji reference arm length is \(L=3.0\times10^9\,\mathrm{m}\), whereas the nominal LISA design adopts \(L=2.5\times10^9\,\mathrm{m}\). These similar design features place Taiji and LISA in the same millihertz observational regime and motivate using a Taiji-like constellation to study how orbital evolution, detector response, and instrumental noise map an SGWB spectrum onto measurable time-delay interferometry (TDI) observables. We implement the present analysis for a Taiji-like constellation. The same spectral-inference strategy can also be adapted to other planned space-based GW detectors, including TianQin \cite{TianQin:2015yph,Hu:2018yqb,TianQin:2020hid} and the DECi-hertz Interferometer Gravitational-wave Observatory (DECIGO) \cite{Seto:2001qf,Isoyama:2018rjb,Kawamura:2020pcg}, once their mission-specific orbital configurations, measurement architectures, detector responses, and noise models are incorporated.

TDI is essential because laser-frequency noise in space-based GW detectors greatly exceeds the displacement fluctuations induced by GWs \cite{LISA:2017pwj,Tinto:2002de,Tinto:2014lxa}. It suppresses this noise by combining appropriately delayed one-way phase measurements between spacecraft \cite{Tinto:2002de,Tinto:2014lxa,Tinto:2003vj}. Under the static equal-arm approximation and the usual symmetry assumptions, the Michelson covariance can be diagonalized by a fixed \(A,E,T\) channel rotation \cite{Prince:2002hp}. For unequal and time-varying arms, however, a single fixed rotation cannot diagonalize the covariance for all segments and frequencies. The response functions, the acceleration noise (ACC) and optical metrology system (OMS) transfer functions all become orbit dependent. For unequal and time-varying arms, the Michelson \(X,Y,Z\) channels are generally correlated. We therefore model their auto- and cross-spectra jointly with a full complex \(3\times3\) covariance matrix for each data segment and frequency. The stochastic-signal and instrumental-noise contributions to this matrix are constructed separately from the detector response functions and noise transfer functions, respectively \cite{Larson:1999we,Robson:2018ifk,Smith:2019wny,Babak:2021mhe,Caprini:2024hue}. To describe the moving constellation, we use second-generation TDI and evaluate the response and noise transfer functions using the orbital configuration associated with each segment \cite{Prince:2002hp,Cornish:2002rt,Vallisneri:2004bn,Petiteau:2008zz,Du:2025xdq}. Relative to the static equal-arm approximation, the resulting spectral deviations are most visible near TDI nulls and at the low- and high-frequency edges of the retained band.

Most SGWB inference studies are formulated in the frequency domain (FD), where detector responses, instrumental-noise power spectral densities (PSDs), and parametric source spectra can be combined efficiently in a Gaussian spectral likelihood \cite{Allen:1997ad,Romano:2016dpx,Smith:2019wny,Babak:2021mhe}. This FD formulation remains statistically convenient and computationally efficient, and we keep it as the inference layer of the present analysis. We use time-domain (TD) simulations because real space-based GW data are finite and sampled time streams that must be combined into delayed TDI observables before spectral analysis. Although an FD treatment provides an efficient description of the ideal spectral response, TD simulations verify that this description remains applicable to orbit-dependent data streams in a realistic mission analysis \cite{Vallisneri:2004bn,Petiteau:2008zz,Du:2025xdq,Baghi:2023qnq,Wu:2025rzi}.

The central objective of this work is therefore to build such a simulation-to-inference framework and to validate the associated detector functions. We generate second-generation Michelson \(X,Y,Z\) TDI time streams for a Taiji-like moving constellation using \trianglesim{} \cite{Du:2025xdq,Du:2025fes}. The simulated streams are divided into finite segments and Fourier transformed. For each segment, the corresponding orbital configuration is used to compute the response functions and noise transfer functions. The retained Fourier coefficients are then analyzed with a segment-wise FD Gaussian likelihood that retains the full complex covariance of the \(X,Y,Z\) channels. In this construction, TD simulations bring the data model closer to the structure of realistic detector outputs, while the FD likelihood preserves the efficiency needed for Bayesian parameter estimation (PE).

The static equal-arm FD configuration provides an idealized baseline for analytic checks and rapid forecasts. The equal-arm TD configuration retains the equal-arm geometry, while the unequal-arm TD configuration further includes time-dependent orbital evolution and unequal-arm second-generation TDI delays. Together, these three detector configurations provide a controlled comparison designed to isolate the effects that enter as the analysis moves toward a more realistic TD setting. Comparing the three cases using full \(XYZ\) covariance likelihoods matched to their respective detector configurations quantifies how explicit TD construction and moving unequal arms modify the response functions, the noise transfer functions, the posterior geometry, and the Bayesian evidence.

This study builds on a broad literature on stochastic-background separation and space-based detector simulation. Previous work established the separation of stochastic signals from instrumental noise and Galactic foregrounds \cite{Adams:2010vc,Adams:2013qma}, quantified compact-binary and phase-transition SGWB components \cite{Karnesis:2021tsh,Caprini:2019egz}, developed flexible spectral-shape reconstruction methods \cite{Caprini:2019pxz,Flauger:2020qyi,Alvey:2023npw,Pozzoli:2023lgz,Pieroni:2020rob}, and studied multi-component separation involving astrophysical backgrounds, cosmological backgrounds, and orbitally modulated Galactic foregrounds \cite{Boileau:2020rpg,Boileau:2021sni}. Bayesian analyses based on TD simulations of space-based GW detectors have demonstrated anisotropic SGWB and Galactic-foreground recovery and have recently extended this direction to multiple isotropic and anisotropic stochastic components \cite{Banagiri:2021ovv,Rieck:2023pej,Criswell:2024hfn,Criswell:2025acn}. Other studies \cite{Muratore:2023gxh,Aimen:2025zxn,Kume:2024xvh,Karnesis:2026gxc,Santini:2025iuj} have examined time-varying unequal arms, second-generation TDI, full multichannel covariance, and flexible instrumental-noise modeling; a related Taiji analysis has also adopted segmented second-generation TDI and full non-diagonal covariance for isotropic SGWB reconstruction \cite{Wang:2020fwa,Wang:2020pkk,Wang:2022sti,Wang:2021njt,Baghi:2023qnq,Jiang:2026lik}. Complementary global-fit work provides the residual-analysis context for stochastic searches \cite{Littenberg:2023xpl,Rosati:2024lcs,DeSanti:2026xhs,Criswell:2026xqk,Deng:2025wgk,Katz:2024oqg}, while TD streams have been used to recover a phase-transition signal in the presence of a modulated Galactic foreground \cite{Hindmarsh:2024ttn}. The present analysis is complementary to these studies, since we test a more crowded multi-component stochastic spectrum by jointly fitting instrumental noise, an effective Galactic foreground, a stochastic astrophysical background, and, in the final benchmark, a cosmological SW component within one full-covariance framework.

In the millihertz band, unresolved Galactic DWD binaries produce a foreground that overlaps with stochastic compact-binary and cosmological backgrounds \cite{Nelemans:2001hp,Robson:2018ifk,Babak:2021mhe}. We neither subtract individually resolvable sources nor construct a fully anisotropic Galactic model \cite{Cornish:2003vj,Timpano:2005gm,Littenberg:2020bxy,Guan:2026qsd,Jiang:2026zto,Mentasti:2023uyi,Digman:2022jmp}. Instead, the unresolved Galactic foreground is represented by an effective isotropic broken power law, and the stochastic astrophysical background is represented by a phenomenological power law; both are inferred together with the instrumental amplitudes in the full \(XYZ\) covariance likelihood \cite{Rosado:2011kv,Farmer:2003pa}. A SW template is then included as a representative cosmological component \cite{Hindmarsh:2013xza,Hindmarsh:2015qta,Hindmarsh:2017gnf,Hindmarsh:2019phv}.

Before applying the detector functions in PE, we first validate the response and noise transfer functions with controlled TD simulations, following the Refs.\cite{Vallisneri:2004bn,Petiteau:2008zz,QuangNam:2022gjz,Du:2025xdq}. The detailed validation setup is presented later; here we emphasize that the simulations are used to check the consistency between the orbit-dependent detector response and the corresponding second-generation TDI data streams generated with \trianglesim{}. We then use Bayesian PE to assess how instrumental noise, astrophysical foregrounds and backgrounds, and cosmological signal parameters can be jointly constrained. Nested sampling (NS) provides posterior samples and Bayesian evidences for model comparison \cite{Skilling:2006gxv,Feroz:2008xx,Feroz:2013hea,Speagle:2019ivv,Ashton:2018jfp,Romero-Shaw:2020owr,Veitch:2014wba,Vallisneri:2007ev}. 

The paper is organized as follows. Section~\ref{sec:tdanalysis} builds the data model by connecting simulated Taiji TDI streams to the FD covariance likelihood used for inference and validates the detector functions used in this construction. Section~\ref{sec:results} applies this framework to foreground and astrophysical-background separation, and the SW search. Section~\ref{sec:conclusion} gives the conclusion of this work.

\section{Simulation-to-Inference Framework}\label{sec:tdanalysis}

This section describes the detector model and data-simulation framework used throughout the analysis. We generate second-generation Michelson TDI strain data in the \(X,Y,Z\) channels, divide the TD streams into segments, and Fourier transform each segment. After restricting the Fourier data to the selected frequency range and excluding bins near the TDI null frequencies, the retained Fourier coefficients are modeled with a segment-dependent complex \(3\times3\) covariance matrix comprising instrumental noise and three distinct stochastic components: the Galactic DWD foreground, the astrophysical background, and the SW spectrum. The FD likelihood retains the full \(XYZ\) covariance to capture correlations induced by unequal, time-dependent arms, while single-component simulations are used to validate the detector response and noise transfer functions.

\subsection{Time-Domain TDI Data and Spectral Components}\label{sec:xyzsources}

The Taiji detector considered here consists of three spacecraft (SCs) forming a nearly equilateral triangular constellation in heliocentric orbit. Each SC carries two movable optical subassemblies (MOSAs). We label the directed inter-spacecraft measurements by ordered pairs \(ij\in\{12,13,23,21,31,32\}\), where \(i\) is the receiving SC and \(j\) is the emitting SC. Following the standard formulation in Ref.~\cite{Du:2025xdq}, the elementary observable \(\eta_{ij}\) is constructed in two steps.

First, one defines the intermediate observable
\begin{equation}
\xi_{ij} = s_{ij} + \frac{\tau_{ij}-\varepsilon_{ij}}{2}
+ \mathbf{D}_{ij}\frac{\tau_{ji}-\varepsilon_{ji}}{2}.
\end{equation}
Here \(s_{ij}\) denotes the science interferometer measurement on the directed link \(j\to i\), \(\varepsilon_{ij}\) denotes the test-mass interferometer measurement, and \(\tau_{ij}\) denotes the reference interferometer measurement.

The time-delay operator \(\mathbf{D}_{ij}\) is defined as
\begin{equation}
\mathbf{D}_{ij} f(t) = f\!\left[t - d_{ij}(t)\right],
\end{equation}
where \(f(t)\) is an arbitrary function, \(d_{ij}(t)\) is the light-travel time along the directed link \(j\to i\), and \(d_{ij}(t)=L_{ij}(t)/c\). Here \(L_{ij}(t)\) is the corresponding time-dependent arm length, while the constant \(L=3.0\times10^9\,\mathrm{m}\) denotes the reference equal-arm length about which the individual \(L_{ij}(t)\) vary.

Next, for each cyclic triple \((i,j,k)\in\{(1,2,3),(2,3,1),(3,1,2)\}\), the final elementary observables are
\begin{equation}
\eta_{ij} = \xi_{ij}
+ \mathbf{D}_{ij} \frac{\tau_{ji} - \tau_{jk}}{2},
\qquad
\eta_{ik} = \xi_{ik}
+ \frac{\tau_{ij} - \tau_{ik}}{2},
\end{equation}
where the first relation defines \(\eta_{12},\eta_{23},\eta_{31}\) and the second defines \(\eta_{13},\eta_{21},\eta_{32}\).

We write the second-generation Michelson TDI channels in the unified delay-polynomial form used by the Taiji simulation convention~\cite{Du:2025xdq}, with standard TDI delay-operator notation~\cite{Tinto:2014lxa}:
\begin{equation}\label{ugl}
d_I(t)=\sum_{ij}\mathbf P_{ij}^{(I)}\eta_{ij}(t),
\qquad I\in\{X,Y,Z\},
\end{equation}
where \(\mathbf{P}_{ij}^{(I)}\) denotes the delay polynomial associated with TDI channel \(I\). We adopt the \((X_2,Y_2,Z_2)\) convention implemented in \trianglesim{}, where the subscript \(2\) indicates second-generation TDI. The \(X_2\) combination is constructed following Ref.~\cite{Du:2025xdq}, while the \(Y_2\) and \(Z_2\) combinations are constructed analogously.

We emphasize that GW signals enter the TDI streams through the science-interferometer terms \(s_{ij}\) at the elementary-link level, before the delay-polynomial combinations are formed. In the detector-response formalism of Ref.~\cite{Smith:2019wny}, a plane-wave perturbation \(h_{ab}(f,\hat n)=\sum\limits_{A = {+,\times}} h_A(f,\hat n)e^A_{ab}(\hat n)\) contributes to an interferometric phase through a contraction between the polarization tensor and the detector tensor, schematically \(e^A_{ab}(\hat n) D^{ab}(f,\hat n)\). Applying the TDI polynomials in Eq.~\eqref{ugl} to these link-level responses gives the channel responses \(\mathcal T_I^{A,\kappa}\) used in Eq.~\eqref{eq:responsematrix}.

Finally, the TD streams are Fourier transformed segment by segment. For each segment and Fourier bin we use the vector
\begin{equation}\label{eq:xyzdatavec}
\tilde{\mathbf d}^{\kappa}(f_k)
=
\left(
\tilde d_X^\kappa(f_k),
\tilde d_Y^\kappa(f_k),
\tilde d_Z^\kappa(f_k)
\right)^{\rm T},
\end{equation}
where \(\kappa\) indexes the data segment and \(f_k\) is the frequency of the \(k\)-th Fourier bin.

Having defined the FD data vector, we now specify how the stochastic GW energy-density spectra and instrumental noise contribute to the spectral matrix of the \(X,Y,Z\) channels. We adopt the standard finite-segment discrete Fourier transform (DFT) normalization, \(2/(T_{\rm seg}f_s^2)\), and place the complex conjugate on the second Fourier coefficient, following Ref.~\cite{Flauger:2020qyi}. For segment \(\kappa\) and frequency bin \(f_k\), the spectrum associated with channels \(I\) and \(J\) is defined as
\begin{equation}\label{eq:xyzcovexpect}
    P_{IJ}^{\kappa}(f_k)
    =
    \frac{2}{T_{\rm seg} f_s^2}
    \left\langle
    \tilde{d}_I^\kappa(f_k)
    \tilde{d}_J^\kappa(f_k)^{*}
    \right\rangle,
    \qquad I,J\in\{X,Y,Z\}.
\end{equation}
Here \(T_{\rm seg}\) is the segment duration and \(f_s\) is the sampling frequency of the TD data. For \(I=J\), \(P_{II}^{\kappa}\) is a real auto-spectral density; for \(I\neq J\), \(P_{IJ}^{\kappa}\) is a generally complex cross-spectral density (CSD) satisfying \(P_{JI}^{\kappa}=(P_{IJ}^{\kappa})^{*}\). We keep the full \(3\times3\) spectral matrix because correlations among the Michelson channels generally persist for unequal, time-dependent arms and cannot be removed exactly by the fixed \(A,E,T\) transformation used in the equal-arm approximation.

For each segment \(\kappa\), we model the spectral matrix as
\begin{equation}\label{eq:xyzmodelmatrix}
    P_{IJ}^{\kappa}(f)
    =
    N_{IJ}^{\kappa}(f)
    +
    S_h(f)\mathcal R_{IJ}^{\kappa}(f),
\end{equation}
where \(N_{IJ}^{\kappa}(f)\) denotes the instrumental-noise contribution and \(\mathcal R_{IJ}^{\kappa}(f)\) is the response function. 
The response function describes how an isotropic stochastic background is mapped into the TDI channels. It includes the TDI transfer functions for the specified arm lengths and is averaged over sky direction and polarization. The instrumental-noise contribution is constructed separately from the ACC and OMS noise spectra~\cite{Smith:2019wny,Babak:2021mhe,Guan:2025idx,Liang:2026wwz}.

Using the one-sided PSD normalization adopted in Ref.~\cite{Guan:2025idx}, the strain PSD \(S_h(f)\) and the dimensionless GW energy density \(\Omega_{\rm GW}(f)\) are related by
\begin{equation}\label{eq:shomega}
    S_h(f)
    =
    \frac{3H_0^2}{4\pi^2}
    \frac{\Omega_{\rm GW}(f)}{f^3},
\end{equation}
where \(H_0\) is the Hubble constant. This scalar strain PSD provides the common input spectrum in Eq.~\eqref{eq:xyzmodelmatrix}; for each channel pair, \(\mathcal R_{IJ}^{\kappa}(f)\) converts it into the corresponding SGWB contribution.

The total stochastic spectrum is modeled as
\begin{equation}\label{eq:omegatotal}
\Omega_{\rm GW}(f)=\Omega_{\rm DWD}(f)+\Omega_{{\rm GW},{\rm ast}}(f)+\Omega_{\rm SW}(f).
\end{equation}
We adopt the same three-component parametrization as Ref.~\cite{Liang:2025zku}, where these spectral components are discussed in detail. For the present analysis, the unresolved Galactic DWD foreground is represented by the effective isotropic broken power law
\begin{equation}\label{eq:dwdmodel}
\Omega_{\rm DWD}(f)=\frac{A_1(f/f_0)^{\alpha_1}}{1+A_2(f/f_0)^{\alpha_2}}, \qquad f_0=16~\mathrm{mHz}.
\end{equation}
The stochastic astrophysical background is modeled as
\begin{equation}\label{eq:astmodel}
\Omega_{{\rm GW},{\rm ast}}(f)=\Omega_{\rm ast}\left(\frac{f}{10^{-3}\,\mathrm{Hz}}\right)^{\varepsilon},
\end{equation}
where \(\Omega_{\rm ast}\) is the amplitude at \(10^{-3}\,\mathrm{Hz}\) and \(\varepsilon\) is the spectral index.

The cosmological component is described by the SW template for a FOPT~\cite{Guan:2025idx,Liang:2025zku},
\begin{equation}\label{eq:swmodel}
\Omega_{\rm SW}(f)=\Omega_0\left(\frac{f}{f_p}\right)^3\left[\frac{7}{4+3(f/f_p)^2}\right]^{7/2},
\end{equation}
where \(\Omega_0\) and \(f_p\) are the peak amplitude and peak frequency, respectively. The sum of these components is converted into \(S_h(f)\) through Eq.~\eqref{eq:shomega}, and the response function \(\mathcal R_{IJ}^{\kappa}(f)\) then determines its contribution to the \(XYZ\) spectral matrix in Eq.~\eqref{eq:xyzmodelmatrix}.

The instrumental-noise model is built from the single-link ACC and OMS spectra. Following the single-link noise convention of Ref.~\cite{Guan:2025idx}, their amplitude spectral densities are written as
\begin{align}
\sqrt{S_{\rm acc}(f)} &= N_{\rm acc}\sqrt{1+\left(\frac{0.4\,{\rm mHz}}{f}\right)^2}\sqrt{1+\left(\frac{f}{8\,{\rm mHz}}\right)^4}\left(\frac{\mathrm{m}}{\mathrm{s}^{2}\sqrt{\mathrm{Hz}}}\right), \label{eq:sacc}\\
\sqrt{S_{\rm OMS}(f)} &= \delta x\sqrt{1+\left(\frac{2\,{\rm mHz}}{f}\right)^4}\left(\frac{\mathrm{m}}{\sqrt{\mathrm{Hz}}}\right). \label{eq:soms}
\end{align}
For the Taiji benchmark, we adopt \(N_{\rm acc}=3\times10^{-15}\) and \(\delta x=8\times10^{-12}\).

These quantities describe the noise in the individual links before the TDI combinations are formed. After the appropriately delayed link measurements are combined into the second-generation Michelson \(X,Y,Z\) channels, the noise contribution is
\begin{equation}\label{eq:noisetransfermatrix}
N_{IJ}^{\kappa}(f)=N_{\rm acc}^{2}\mathcal N^{\rm acc,\kappa}_{IJ}(f)+\delta x^{2}\mathcal N^{\rm OMS,\kappa}_{IJ}(f), \qquad I,J\in\{X,Y,Z\}.
\end{equation}
Here \(\mathcal N^{\rm acc,\kappa}_{IJ}(f)\) and \(\mathcal N^{\rm OMS,\kappa}_{IJ}(f)\) are the post-TDI noise transfer functions for the full \(XYZ\) spectral matrix. Expanding the delay operators yields explicit link-polynomial expressions, and the corresponding noise transfer functions for Michelson PSDs and CSDs are derived in Ref.~\cite{QuangNam:2022gjz}.

The SGWB contribution is described by the response functions \(\mathcal R_{IJ}^{\kappa}(f)\). Adopting the standard convention for the response of space-based GW detectors~\cite{Cornish:2002rt,Smith:2019wny,Babak:2021mhe,Guan:2025idx}, the response to an isotropic, unpolarized SGWB is
\begin{equation}\label{eq:responsematrix}
\mathcal R_{IJ}^{\kappa}(f)=\frac{1}{2}\sum_{A=+,\times}\left\langle\mathcal T_I^{A,\kappa}(f,\hat n)\mathcal T_J^{A,\kappa}(f,\hat n)^*\right\rangle_{\hat n}.
\end{equation}
Here \(\mathcal T_I^{A,\kappa}(f,\hat n)\) denotes the TDI response of channel \(I\) to a unit-amplitude plane GW at Fourier frequency \(f\), propagating in direction \(\hat n\) with tensor polarization \(A=+,\times\). The brackets denote the sky average, \(\langle X\rangle_{\hat n}\equiv(4\pi)^{-1}\int d^2\hat n\,X\) and the factor \(1/2\) averages over the two tensor polarizations.

Together, these definitions specify the segment-dependent spectral model in Eq.~\eqref{eq:xyzmodelmatrix}. The segment dependence is required because each one-day segment corresponds to a different orbital configuration. Accordingly, both the response functions and the noise transfer functions are evaluated using the orbital information associated with the corresponding segment.

\subsection{Validation of Response and Noise Transfer Functions}\label{sec:tdsimvalidation}

The validation runs and the subsequent PE analyses use a common TD data-generation and FD spectral-analysis setup. For each 360-day simulated data set, \trianglesim{} generates second-generation Michelson TDI streams~\cite{Tinto:2003vj,Wang:2020pkk} over $T_{\rm obs}=360$ days at a sampling frequency $f_s=0.4\,\mathrm{Hz}$. The streams are divided into $N_{\rm seg}=360$ one-day segments of duration $T_{\rm seg}=86400\,\mathrm{s}$. For each segment, we apply a Hann window to the TD samples of the \(X,Y,Z\) channels and compute their Fourier transforms. The resulting Fourier coefficients are then used to construct the one-sided \(XYZ\) auto- and cross-spectral estimates with the normalization defined in Eq.~\eqref{eq:xyzcovexpect}. The positive-frequency grid extends to the Nyquist frequency, $0.2\,\mathrm{Hz}$. Both the validation diagnostics and the PE analyses use the same outer frequency range, $10^{-4}\,\mathrm{Hz}\leq f\leq0.15\,\mathrm{Hz}$. For the PE likelihood, we also excise the portions of the intervals $[f_m^{\rm null}-\Delta f_{\rm notch}^{\rm half},f_m^{\rm null}+\Delta f_{\rm notch}^{\rm half}]$. Here, $f_m^{\rm null}=mc/(4L)$, with $m=1,2,3,\ldots$, denotes the nominal equal-arm null frequencies of the second-generation Michelson observable~\cite{Wang:2024alm}, and we adopt $\Delta f_{\rm notch}^{\rm half}=0.010\,\mathrm{Hz}$. To reveal the null structure and the associated rapid spectral variation, the diagnostic figures retain and display all bins within these neighborhoods, which are shaded gray.

Numerical second-generation TDI for realistic, time-varying unequal-arm LISA- and Taiji-like constellations is already well developed, including orbit-based TDI simulations, unequal-arm sensitivity studies, and integrated simulation-and-analysis frameworks~\cite{Wang:2017aqq,Wang:2020fwa,Wang:2020pkk,Wang:2024ssp,Du:2025xdq}. These works provide the numerical and methodological foundations for the present setup. Here, we validate the detector model through a component-wise comparison with simulated second-generation TDI streams. We compare the spectra predicted by the segment-dependent ACC and OMS noise transfer functions and response functions with the corresponding \(XYZ\) auto- and cross-spectral estimates from the simulations.

Rather than validating only the combined spectrum~\cite{QuangNam:2022gjz,Baghi:2023qnq}, we exploit the linear decomposition in Eqs.~\eqref{eq:xyzmodelmatrix} and~\eqref{eq:noisetransfermatrix} to test separately the three additive contributions entering the detector model: the ACC noise, the OMS noise, and the SGWB response. We therefore perform three component-isolated runs. In the ACC-only run, only the ACC link-noise PSD is enabled, while the OMS link-noise PSD and the SGWB injection are set to zero. In the OMS-only run, only the OMS link-noise PSD is enabled, while the ACC link-noise PSD and the SGWB injection are set to zero. In the response-only run, both instrumental-noise PSDs are set to zero, and a unit strain PSD, \(S_h(f)=1\,\mathrm{Hz}^{-1}\), is injected into every simulated frequency bin. In the last case, the instrumental-noise contribution in Eq.~\eqref{eq:xyzmodelmatrix} vanishes, giving
\[
P_{IJ}^{\kappa}(f)=\mathcal R_{IJ}^{\kappa}(f)\,\mathrm{Hz}^{-1}
\]
under the adopted one-sided PSD convention. The realization-averaged CSD therefore directly estimates the response function for each channel pair \(IJ\). 

The direct simulation-to-calculation comparison is carried out for the first one-day segment of the unequal-arm orbit. For each of the three component-isolated cases, we generate \(N_{\rm real}=1000\) statistically independent realizations with \trianglesim{} and average their spectra. We evaluate all independent entries of the Hermitian \(XYZ\) spectral matrix: \(XX\), \(YY\), \(ZZ\), \(XY\), \(XZ\), and \(YZ\). The diagonal entries are treated as real auto-spectra, whereas the three independent off-diagonal entries are retained as complex cross-spectra and assessed through both their amplitudes and phases.

For realization \(n\), segment \(\kappa\), and retained frequency bin \(f_k\), the single-realization cross-spectral estimator is
\begin{equation}\label{eq:observedcsd}
    \widehat G_{IJ,n}^{\kappa}(f_k)
    =
    \frac{2}{T_{\rm seg} f_s^2}
    \tilde d_{I,n}^{\kappa}(f_k)
    \tilde d_{J,n}^{\kappa}(f_k)^{*},
    \qquad I,J\in\{X,Y,Z\}.
\end{equation}
Here \(\tilde d_{I,n}^{\kappa}(f_k)\) is the Fourier coefficient of the data. This estimator yields the three real auto-spectral estimates and the three independent complex cross-spectral estimates, including their phases, for each component-isolated run. Appendix~\ref{app:detector-validation} presents a detailed, component-wise comparison of the realization-averaged estimates with the corresponding calculated ACC and OMS noise transfer functions and response functions.

Even when the calculated response or noise transfer function is exactly consistent with the simulated model, the average of a finite number of realizations will not coincide exactly with it in every frequency bin. The simulation-to-calculation residuals must therefore be interpreted relative to the statistical scatter expected for the \(N_{\rm real}\)-realization average. The covariance of these complex spectral estimators follows the standard covariance relations for multichannel complex-periodogram estimators used in SGWB analyses~\cite{Romano:2016dpx,Baghi:2023qnq,Hartwig:2023pft}. Here, we use this covariance to predict the finite-realization variance of the averaged estimator \(\widehat G_{IJ}^{\kappa}(f_k)\) around the calculated spectral element \(P_{IJ}^{\kappa}(f_k)\). This prediction sets the reference scale for the normalized residuals introduced below.

For the channel pair $IJ$, let $P_{IJ}^{\kappa}(f_k)$ denote the calculated auto- or cross-spectral element for the component being tested. For the estimator in Eq.~\eqref{eq:observedcsd}, the ensemble mean and finite-realization average are
\begin{equation}\label{eq:realavgcsd}
    \left\langle \widehat G_{IJ,n}^{\kappa}(f_k)\right\rangle
    =
    P_{IJ}^{\kappa}(f_k),
    \qquad
    \widehat G_{IJ}^{\kappa}(f_k)
    =
    \frac{1}{N_{\rm real}}
    \sum_{n=1}^{N_{\rm real}}
    \widehat G_{IJ,n}^{\kappa}(f_k).
\end{equation}
We quantify the residual of simulation-to-calculation difference
\begin{equation}
    r_{IJ}^{\kappa}(f_k)
    =
    \frac{
    \left|\widehat G_{IJ}^{\kappa}(f_k)-P_{IJ}^{\kappa}(f_k)\right|
    }{
    \left|P_{IJ}^{\kappa}(f_k)\right|
    },
    \label{eq:crossspectrum_relative_residual}
\end{equation}
and use the coherence between different channels ~ \cite{Hartwig:2023pft},
\begin{equation}
    \rho^{\kappa}_{IJ}(f_k)
    =
    \frac{\left|P^{\kappa}_{IJ}(f_k)\right|}
    {\sqrt{P^{\kappa}_{II}(f_k)P^{\kappa}_{JJ}(f_k)}},
    \qquad
    0\leq \rho^{\kappa}_{IJ}\leq 1 .
    \label{eq:spectral_element_coherence}
\end{equation}
The coherence measures how large the complex channel-pair element is relative to the two associated auto-spectra. For nonzero $P_{IJ}^{\kappa}(f_k)$, the complex Gaussian spectral covariance derived in Appendix~\ref{app:rij-derivation} gives the finite-realization expectation
\begin{equation}
    \left\langle
    \left[r_{IJ}^{\kappa}(f_k)\right]^2
    \right\rangle
    =
    \frac{1}
    {N_{\rm real}\left[\rho_{IJ}^{\kappa}(f_k)\right]^2}.
    \label{eq:relative_residual_square_expectation}
\end{equation}

\begin{figure}[t]
    \centering
    \makebox[\linewidth][c]{\includegraphics[width=1.12\linewidth]{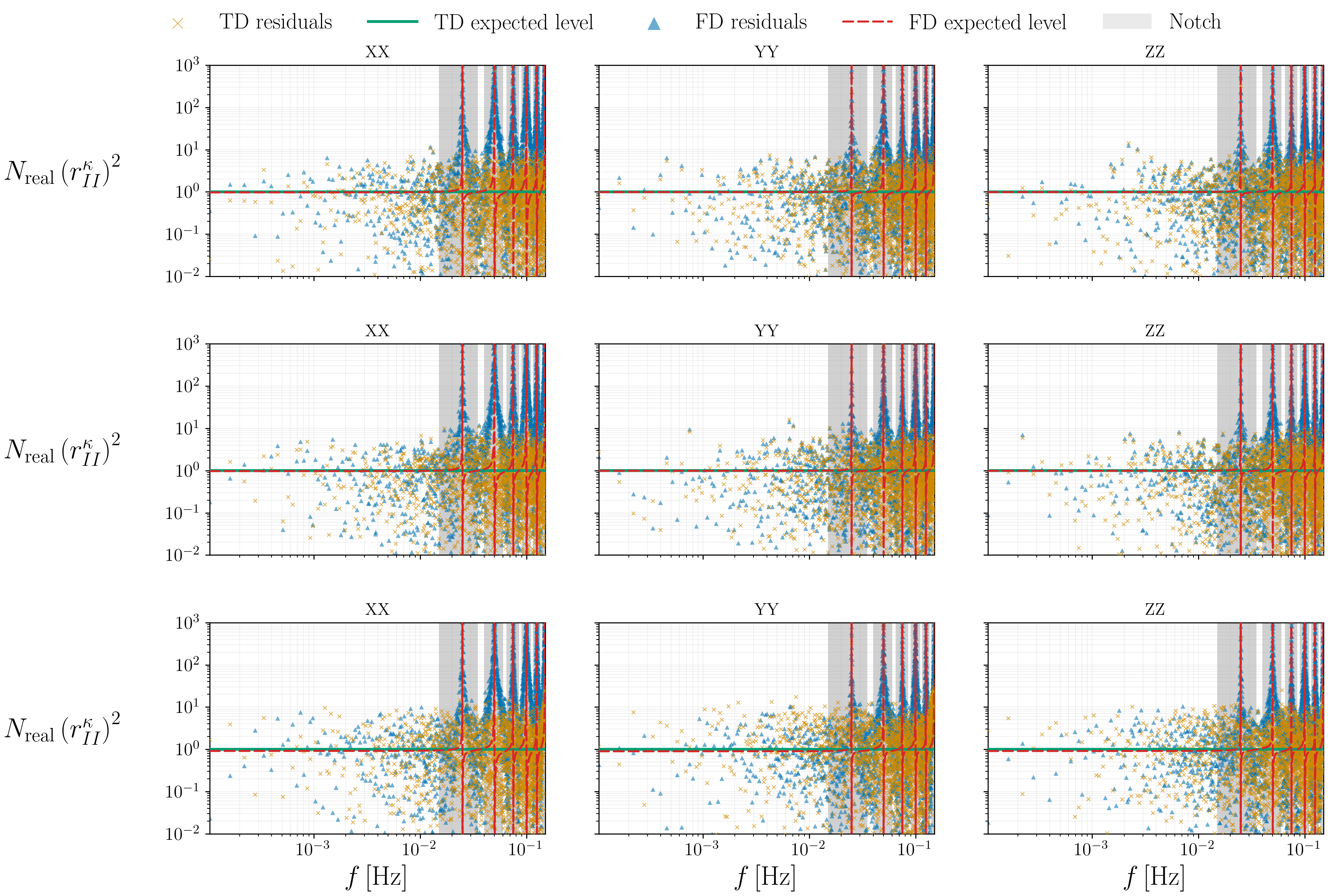}}
    \caption{
    Scaled squared normalized residuals for the diagonal validation entries. Each panel shows $N_{\rm real}[r_{II}^{\kappa}(f_k)]^2$ for the mean of 1000 \trianglesim{} realizations. Orange crosses compare this realization mean with the calculated unequal-arm TD function for the same one-day segment. Blue triangles compare the same unequal-arm realization mean with the static equal-arm FD function on the identical frequency grid. The green solid curve is the matched unequal-arm TD expectation from Eq.~\eqref{eq:relative_residual_square_expectation} multiplied by $N_{\rm real}$; the red dashed curve is the corresponding scaled reference constructed from the static equal-arm FD functions. Because the blue points do not come from an independent static-FD simulation, their departures can contain both finite-realization scatter and deterministic differences between the two configurations. Pale gray bands mark the TDI-null neighborhoods displayed in the diagnostics but excluded from the PE likelihood. Rows correspond to the ACC noise transfer functions, OMS noise transfer functions, and response functions, and columns correspond to $XX$, $YY$, and $ZZ$.
    }
    \label{fig:diagonaluncertaintyscatter}
\end{figure}

\begin{figure}[t]
    \centering
    \makebox[\linewidth][c]{\includegraphics[width=1.12\linewidth]{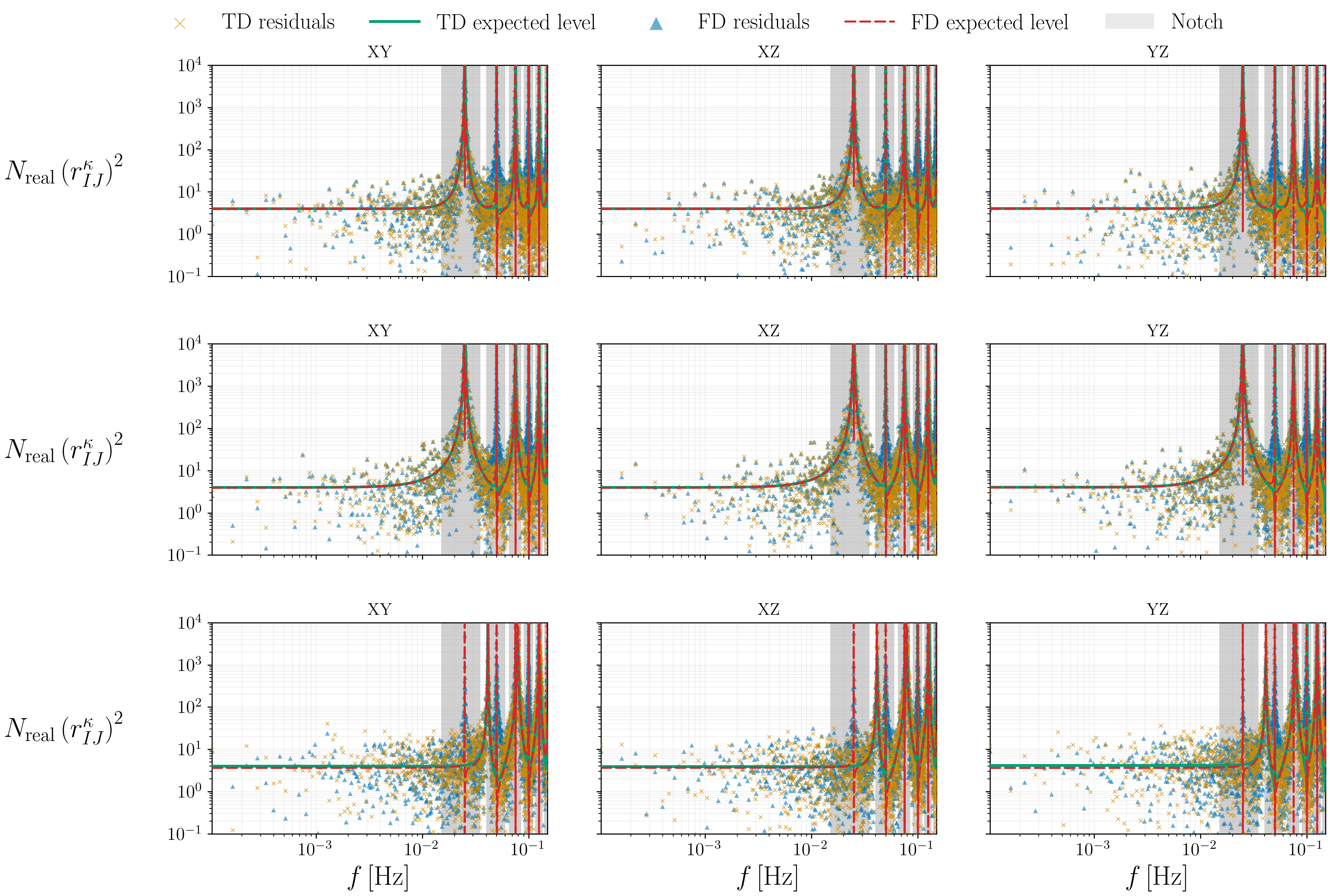}}
    \caption{
    Scaled squared normalized residuals for the independent off-diagonal entries $XY$, $XZ$, and $YZ$, using the same realizations, model references, plotting conventions, and TDI-null bands as Fig.~\ref{fig:diagonaluncertaintyscatter}. The three rows again correspond to the ACC noise transfer functions, OMS noise transfer functions, and response functions. Because the off-diagonal coherence $\rho_{IJ}^{\kappa}$ can be substantially below unity, $N_{\rm real}$ times Eq.~\eqref{eq:relative_residual_square_expectation} predicts a higher matched-model residual level than for the diagonal entries. Away from the null neighborhoods, the unequal-arm TD residuals follow this coherence-dependent finite-realization scale. The largest excursions occur where the relevant cross-spectral element becomes small and rapidly varying.
    }
    \label{fig:crosscoherencescatter}
\end{figure}

The curves overlaid in Figs.~\ref{fig:diagonaluncertaintyscatter} and~\ref{fig:crosscoherencescatter} show the predictions of Eq.~\eqref{eq:relative_residual_square_expectation} multiplied by \(N_{\rm real}\), and therefore give the expected values of the plotted \(N_{\rm real}(r_{IJ}^{\kappa})^2\) under the corresponding spectral functions. Away from the shaded TDI-null neighborhoods and the frequency-band edges, the unequal-arm TD points in Figs.~\ref{fig:diagonaluncertaintyscatter} and~\ref{fig:crosscoherencescatter} are broadly consistent with the expectation \(N_{\rm real}\langle(r_{IJ}^{\kappa})^2\rangle\) obtained from Eq.~\eqref{eq:relative_residual_square_expectation}. For a nonzero auto-spectrum, \(\rho_{II}=1\), so the expectation of the scaled squared residual is unity. With \(N_{\rm real}=1000\), this corresponds to a relative uncertainty of approximately \(1/\sqrt{N_{\rm real}}\simeq3\%\), even when the simulated spectra and calculated functions are statistically consistent. For an off-diagonal cross-spectrum, \(\rho_{IJ}\) can be smaller than unity. Consequently, the off-diagonal panels can exhibit larger residuals. This effect is strongest where \(\rho_{IJ}\) approaches zero, around which the normalized residual is ill-conditioned and the phase estimate becomes unstable.

TDI nulls are a known feature of Michelson-type observables. Recent work has shown that alternative second-generation TDI configurations with fewer characteristic nulls can improve spectral stability, noise characterization, segment-boundary behavior, and high-frequency FD analysis~\cite{Wang:2024alm,Wang:2024hgv,Wang:2025mee,Wang:2025voa}. In the present analysis, the corresponding null neighborhoods serve as sensitive diagnostic regions: near the TDI nulls and the frequency-band edges, small spectral denominators and rapidly varying cross-spectral phases make the comparison especially sensitive to local mismatches. This behavior motivates the common notch mask and band cuts adopted in the subsequent likelihood analysis.

The differences among the three configurations are most pronounced in the vicinity of the nominal TDI nulls. In the static equal-arm limit, exact cancellation of the delay-polynomial factors at discrete frequencies can drive individual auto- and cross-spectral elements to zero. For unequal, time-dependent arms, orbital variations in the arm lengths and link directions perturb the cancellation conditions, leaving the corresponding elements small but finite. This can improve the local conditioning of the full $XYZ$ covariance. The spectra nevertheless remain strongly suppressed and rapidly varying near the nulls, and we therefore apply the same notch mask in all three analyses.

Taken together, the agreement between the simulated spectra and the calculated detector functions demonstrates the internal consistency of the end-to-end framework. This comparison is not fully independent, however, because \trianglesim{} is used both to generate the TD streams and to evaluate the orbit-dependent response and noise transfer functions.

\subsection{Full-Covariance Bayesian Inference}\label{sec:inferencesetup}

PE is performed with a complex Gaussian likelihood in FD for the retained Fourier coefficients in the $XYZ$ basis. The TD simulations supply the data streams and diagnostic products, while the likelihood, Fisher forecasts, and signal-to-noise ratio (SNR) diagnostics are evaluated using the spectral functions defined above. Posterior samples and evidence values are obtained with NS using \pkg{Bilby.dynesty}~\cite{Skilling:2006gxv,Ashton:2018jfp,Romero-Shaw:2020owr,Speagle:2019ivv}.

For each retained segment and frequency, the Fourier coefficient vector in Eq.~\eqref{eq:xyzdatavec} is modeled as a circular complex Gaussian variable in the original $XYZ$ basis. The covariance of this Fourier vector is obtained from the same PSD matrix introduced in Sec.~\ref{sec:xyzsources},
\begin{equation}\label{eq:xyzcov}
\mathbf{C}^{\kappa}(f_k)
=
\frac{T_{\rm seg} f_s^2}{2}
\mathbf{P}^{\kappa}(f_k).
\end{equation}
Equation~\eqref{eq:xyzcov} only converts the one-sided PSD matrix \(\mathbf P^\kappa\) into the covariance of the retained Fourier coefficients. Explicitly, in the \(X,Y,Z\) channel ordering,
\[
\mathbf P^\kappa(f_k)
=
\begin{pmatrix}
P_{XX}^\kappa(f_k) & P_{XY}^\kappa(f_k) & P_{XZ}^\kappa(f_k)\\
P_{YX}^\kappa(f_k) & P_{YY}^\kappa(f_k) & P_{YZ}^\kappa(f_k)\\
P_{ZX}^\kappa(f_k) & P_{ZY}^\kappa(f_k) & P_{ZZ}^\kappa(f_k)
\end{pmatrix},
\]
where \(\mathbf P^\kappa(f_k)\) is Hermitian. The corresponding Whittle likelihood is~\cite{Franciolini:2025leq,Liu_2024}
\begin{equation}\label{eq:xyzlikelihood}
\ln\mathcal{L}_{XYZ}
=
-\sum_{\kappa,k}
\left[
\ln\det\!\left(\pi\mathbf C^\kappa(f_k)\right)
+
(\tilde{\mathbf d}^{\kappa}(f_k))^\dagger
(\mathbf{C}^{\kappa}(f_k))^{-1}
\tilde{\mathbf d}^{\kappa}(f_k)
\right].
\end{equation}
The determinant and inverse are taken over the three Michelson channels. Thus each retained bin contributes the full covariance of the three auto spectra and three independent complex off-diagonal spectra, rather than a product of scalar channel likelihoods.

For the PE runs we use the parameter set
\begin{equation}
\boldsymbol{\theta}
=
\left(
N_{\rm acc},\delta x,\log_{10}A_1,\alpha_1,\log_{10}A_2,\alpha_2,
\log_{10}\Omega_{\rm ast},\varepsilon,\log_{10}\Omega_0,
\log_{10}(f_p/\mathrm{Hz})
\right).
\end{equation}
This list defines the full parameter pool used in the paper; individual benchmarks activate only the relevant subset, as specified below and in Table~\ref{tab:pepriors}. For a chosen fiducial point, we write the local Fisher information matrix (FIM) as~\cite{Tegmark:1996bz,Adams:2010vc,Contaldi:2020rht}
\begin{equation}\label{eq:fim}
    F_{ij}
    =
    \sum_{\kappa,k}
    {\rm Tr}\!\left[
    (\mathbf P^\kappa(f_k))^{-1}
    \frac{\partial \mathbf P^\kappa(f_k)}{\partial \theta_i}
    (\mathbf P^\kappa(f_k))^{-1}
    \frac{\partial \mathbf P^\kappa(f_k)}{\partial \theta_j}
    \right],
\end{equation}
with derivatives taken with respect to the sampled parameters. Here the FIM is used only for a local Gaussian forecast, and later the posterior samples and evidences come from the full NS runs.

For the SNR diagnostics we use the final full-covariance expressions
\begin{equation}
\begin{aligned}
\mathrm{SNR}_{\rm abs}^2
&=
\sum_{\kappa,k}
\operatorname{Tr}\!\left[
\left(\mathbf P_{\rm noise}^{\kappa}(f_k)\right)^{-1}
\mathbf P_{\rm targ}^{\kappa}(f_k)
\left(\mathbf P_{\rm noise}^{\kappa}(f_k)\right)^{-1}
\mathbf P_{\rm targ}^{\kappa}(f_k)
\right],
\\
\mathrm{SNR}_{\rm rel}^2
&=
\sum_{\kappa,k}
\operatorname{Tr}\!\left[
\left(\mathbf P_{\rm base}^{\kappa}(f_k)\right)^{-1}
\mathbf P_{\rm targ}^{\kappa}(f_k)
\left(\mathbf P_{\rm base}^{\kappa}(f_k)\right)^{-1}
\mathbf P_{\rm targ}^{\kappa}(f_k)
\right].
\end{aligned}
\label{eq:snrdefs}
\end{equation}
where \(\mathbf P_{\rm targ}\) is the full PSD matrix of the component whose detectability is being quoted, and \(\mathbf P_{\rm noise}\) is the instrumental-noise PSD matrix. The absolute SNR measures the target against instrumental noise alone, while the relative SNR measures it against the baseline covariance \(\mathbf P_{\rm base}\) used in the corresponding search. A compact derivation is given in Appendix~\ref{app:snr-derivation}.

The two SNR definitions are introduced solely to preserve the same detectability convention as the standard single-channel SNR expression \cite{Liang:2025zku}. In Sec.~\ref{sec:astrofgbg}, \(\mathbf P_{\rm targ}\) denotes the stochastic astrophysical background, and \(\mathbf P_{\rm base}\) denotes the sum of instrumental noise and the unresolved Galactic foreground. In Sec.~\ref{sec:swsrch}, \(\mathbf P_{\rm targ}\) denotes the SW contribution, while \(\mathbf P_{\rm base}\) includes instrumental noise, the unresolved Galactic foreground, and the stochastic astrophysical background.

\begin{table}[t]
    \centering
    \small
    \setlength{\tabcolsep}{4pt}
    \renewcommand{\arraystretch}{1.15}
    \begin{tabularx}{\linewidth}{@{} l >{\centering\arraybackslash}X l @{}}
    \toprule
    Parameter & Benchmark & Uniform prior \\
    \midrule
    $N_{\rm acc}/10^{-15}$ & $3.0$ & $[0,\,20]$ \\
    $\delta x/10^{-12}$ & $8.0$ & $[0,\,20]$ \\
    $\log_{10} A_1$ & $-15.40$ & $[-18,\,-12]$ \\
    $\alpha_1$ & $-5.7$ & $[-8,\,-2]$ \\
    $\log_{10} A_2$ & $-6.32$ & $[-10,\,-2]$ \\
    $\alpha_2$ & $-6.2$ & $[-8,\,-2]$ \\
    $\log_{10}\Omega_{\rm ast}$ & $-11.5$ & $[-14,\,-9]$ \\
    $\varepsilon$ & $2/3$ & $[-1,\,3]$ \\
    $\log_{10}\Omega_0$ & $-11.5$ & $[-13,\,-10]$ \\
    $\log_{10}(f_p/\mathrm{Hz})$ & $-2.25$ & $[-3,\,-0.5]$ \\
    \bottomrule
    \end{tabularx}
    \caption{Prior ranges and benchmark values used in the PE runs. The uniform-prior column gives the sampled ranges adopted in the analyses, while the benchmark column lists the fiducial values.}
    \label{tab:pepriors}
\end{table}

The PE runs evaluate Eq.~\eqref{eq:xyzlikelihood} using uniform priors. Table~\ref{tab:pepriors} lists the prior ranges and fiducial values used in the analysis; for each benchmark, only the subset of parameters relevant to the corresponding model is included.

Throughout the results section, we report relative parameter uncertainties as percentages. For a parameter \(\theta\) with injected value \(\theta_{\rm inj}\), we define
\begin{equation}
    \delta\theta
    \equiv
    \frac{\Delta \theta}{|\theta_{\rm inj}|},
\label{eq:relative_uncertainty}
\end{equation}
where \(\Delta \theta\) is the uncertainty estimated either from the inverse FIM or from the corresponding posterior samples.

For model comparison, each NS run returns a Bayesian evidence \(\mathcal Z\), defined as the likelihood marginalized over all active model parameters under their adopted priors. We report the evidence ratio through the Bayes factor (BF),
\[
\ln \mathrm{BF}
\equiv
\ln\left(\frac{\mathcal Z_{\rm alt}}{\mathcal Z_{\rm null}}\right),
\]
where \(\mathcal Z_{\rm alt}\) is the evidence for the alternative model containing the additional target component, while \(\mathcal Z_{\rm null}\) is the evidence for the corresponding baseline model with that component omitted and all common components retained. Following the Bayesian model-selection scale used in Ref.~\cite{Trotta:2008qt}, we interpret \(\ln \mathrm{BF}\) values of order unity as modest support and \(\ln \mathrm{BF}\gtrsim 8\) as strong evidence for the extended model.

\section{Inference Results}\label{sec:results}

We now apply the simulation and inference framework defined in Sec.~\ref{sec:tdanalysis} to a sequence of tests with increasing spectral complexity. The goal of this section is to assess the performance of the full \(3\times3\) covariance of the \(XYZ\) channels as the analysis progresses from a controlled recovery problem, to a joint foreground and stochastic-background inference problem, and finally to an SW-background search.

Throughout the science benchmarks, we compare three matched analysis configurations: a static equal-arm FD baseline, an equal-arm TD simulated-data analysis, and an unequal-arm TD simulated-data analysis. These configurations investigate the combined impact of explicit TD data construction and the time-dependent unequal-arm response functions and noise transfer functions, using a full-covariance \(XYZ\) likelihood matched to each configuration.

\begin{figure}[t]
    \centering
    \includegraphics[width=0.84\linewidth]{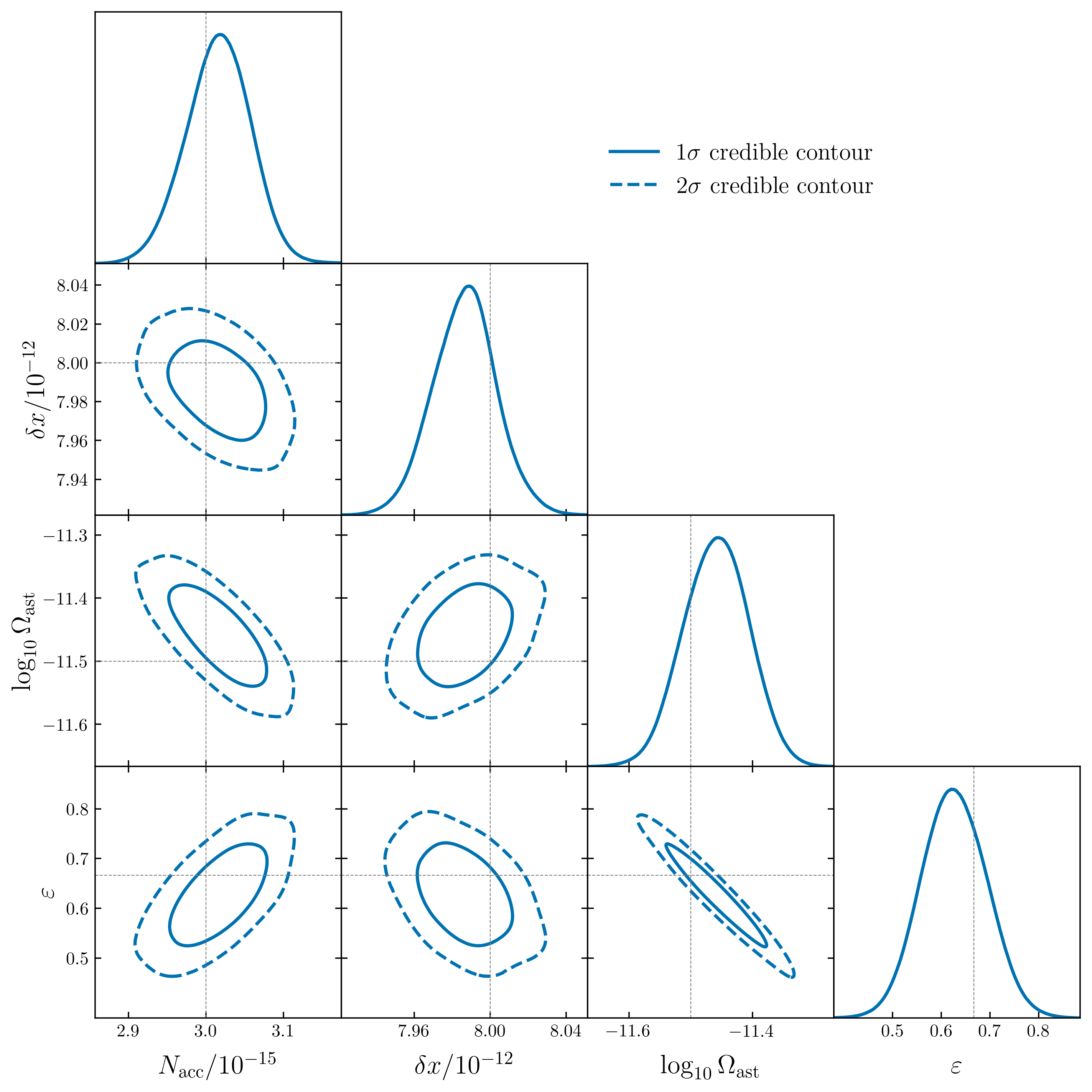}
    \caption{Representative posterior for the minimal four-parameter recovery check with full covariance in the $XYZ$ basis. The fit uses the shorter ten-day unequal-arm TD data set described above, with active parameters $(N_{\rm acc},\delta x,\log_{10}\Omega_{\rm ast},\varepsilon)$. Solid and dashed blue lines indicate the $1\sigma$ and $2\sigma$ credible contours, respectively, and gray crosshairs mark the injected fiducial values. The injected values lie within the displayed credible contours. The dominant degeneracy directions follow the expected spectral structure: $N_{\rm acc}$ and $\delta x$ trade through their joint contribution to the instrumental PSD, while $\log_{10}\Omega_{\rm ast}$ and $\varepsilon$ trade through the amplitude and slope of the power-law background. The corresponding numerical recovery is listed in Table~\ref{tab:xyzpreview}.}
    \label{fig:xyzpevalidation}
\end{figure}

\subsection{Validation of Full-Covariance Parameter Estimation}\label{sec:fourparam}

Before turning to the foreground plus astrophysical background and SW analyses, we test the Bayesian likelihood with a minimal four-parameter benchmark. The direct simulation, transfer function, and response function checks have already been performed in Sec.~\ref{sec:tdsimvalidation}; here the purpose is to verify that the same spectral functions, with full covariance in the $XYZ$ basis, can be used in a PE run. For this recovery test, we use a shorter ten-day data set, divided into ten one-day segments. All subsequent benchmark analyses use the full one-year data sets defined in Sec.~\ref{sec:tdsimvalidation}. We set both the Galactic foreground and the SW contribution to zero, and fit the parameter vector
\begin{equation}\label{wuu}
    (N_{\rm acc},\delta x,\log_{10}\Omega_{\rm ast},\varepsilon).
\end{equation}
Only these four entries of Table~\ref{tab:pepriors} are included in this controlled recovery test.

Fig.~\ref{fig:xyzpevalidation} shows the joint and marginalized posterior distributions obtained from the four-parameter analysis of a single TD realization. The corresponding posterior means, reported in Table~\ref{tab:xyzpreview}, are consistent with the injected values for all four parameters. 

\begin{figure}[t]
    \centering
    \includegraphics[width=0.84\linewidth]{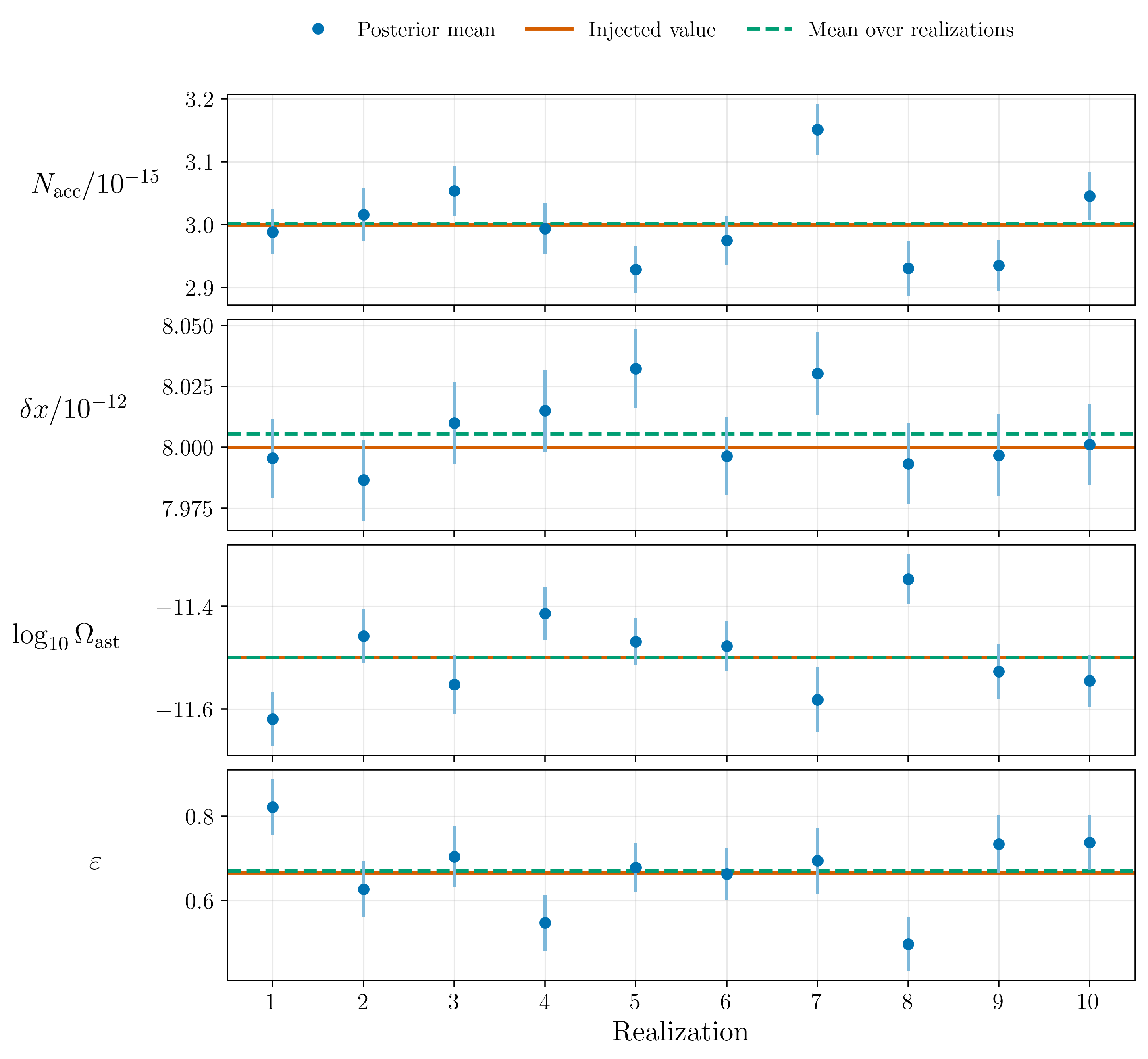}
    \caption{Ten-realization recovery summary for the minimal validation benchmark with full covariance in the $XYZ$ basis. Each point is the posterior mean from one independent ten-day TD realization, and the vertical bar is the corresponding posterior standard deviation. Orange horizontal lines indicate the injected parameter values, while green dashed lines show the averages of the posterior means across the ten realizations.}
    \label{fig:xyzensemblevalidation}
\end{figure}

To assess variation across data realizations, we repeat this validation using ten independent TD realizations. Fig.~\ref{fig:xyzensemblevalidation} shows the posterior mean and posterior standard deviation of each parameter for every realization, allowing the scatter between realizations to be compared directly with the inferred posterior uncertainties.

\begin{table}[t]
    \centering
    \small
    \setlength{\tabcolsep}{6pt}
    \renewcommand{\arraystretch}{1.12}
    \begin{tabular}{lccc}
    \toprule
    Parameter & Injected & Posterior mean & Relative uncertainty \\
    \midrule
    $N_{\rm acc}/10^{-15}$ & $3.000$ & $3.016$ & $1.38\%$ \\
    $\delta x/10^{-12}$ & $8.000$ & $7.987$ & $0.21\%$ \\
    $\log_{10}\Omega_{\rm ast}$ & $-11.500$ & $-11.459$ & $0.45\%$ \\
    $\varepsilon$ & $0.667$ & $0.627$ & $9.98\%$ \\
    \bottomrule
    \end{tabular}
    \caption{Compact recovery summary for the representative ten-day, ten-segment validation benchmark with full covariance in the $XYZ$ basis, shown in Fig.~\ref{fig:xyzpevalidation}. Relative uncertainties follow Eq.~\eqref{eq:relative_uncertainty}.}
    \label{tab:xyzpreview}
\end{table}

The dashed green lines show the mean over the ten recovered posterior means. In the parameter order of Eq.\eqref{wuu}, the injected vector is $(3.0,8.0,-11.5,\frac{2}{3})$, while the average recovered posterior mean is $(3.0018,8.0057,-11.4995,0.6707)$. The realization scatter is therefore visible but remains modest for the purpose of validating the full covariance likelihood in the $XYZ$ basis.

Together with the checks of noise transfer functions and response functions in Sec.~\ref{sec:tdsimvalidation}, these inference tests validate the spectral functions and the full-covariance \(XYZ\) likelihood at the PE level. In the ten-realization validation, we perform the same recovery test on each realization, record the posterior mean of each fitted parameter, and compare the average of these recovered means with the injected values. Overall, the averaged posterior means remain close to the injections, indicating that the pipeline provides an approximately unbiased parameter recovery in this controlled benchmark and can therefore be used for the subsequent PE analyses. The most visible residual offset appears in the OMS noise parameter. This is plausibly explained by the high-frequency notching and cutoff applied in the analysis: these operations remove a substantial fraction of the frequency range where the OMS contribution is most pronounced, and therefore weaken the information available for recovering the OMS noise parameter. Nevertheless, the absolute size of the recovered OMS offset remains acceptable for this validation test and does not indicate a significant bias in the full pipeline.

\subsection{Detection of Astrophysical Background}\label{sec:astrofgbg}

In this section, we consider a more complex spectrum. The target signal is the stochastic astrophysical background, while the unresolved Galactic DWD foreground and the instrumental noise amplitudes are inferred simultaneously. The sampled parameter vector is
\begin{equation}
    \boldsymbol{\theta}_{\rm bg+fg}
    =
    \left(
    N_{\rm acc},
    \delta x,
    \log_{10}A_1,
    \alpha_1,
    \log_{10}A_2,
    \alpha_2,
    \log_{10}\Omega_{\rm ast},
    \varepsilon
    \right).
\end{equation}
The eight-parameter analysis extends the preceding four-parameter analysis by jointly inferring four additional parameters that describe the DWD foreground. The central question is whether the amplitude and spectral index of the astrophysical background remain identifiable after marginalizing over this flexible effective foreground model. We compare three internally matched simulation-to-inference configurations. In the equal-arm FD reference, both the injected spectra and the likelihood are constructed with the static equal-arm FD covariance. In the equal-arm TD analysis, TD simulated streams are analyzed with the corresponding equal-arm TD spectral functions. In the unequal-arm TD analysis, the data are generated using the time-dependent unequal-arm orbit, and the likelihood uses the orbit-dependent response and noise transfer functions, evaluated separately in each segment. Thus, in the matched comparison, varying-arm simulated data are never fitted with a static equal-arm covariance; that deliberately inconsistent case is reserved for the diagnostic mismatch test below.

\begin{figure}[t!]
    \centering
    \makebox[\linewidth][c]{\includegraphics[width=1.10\linewidth]{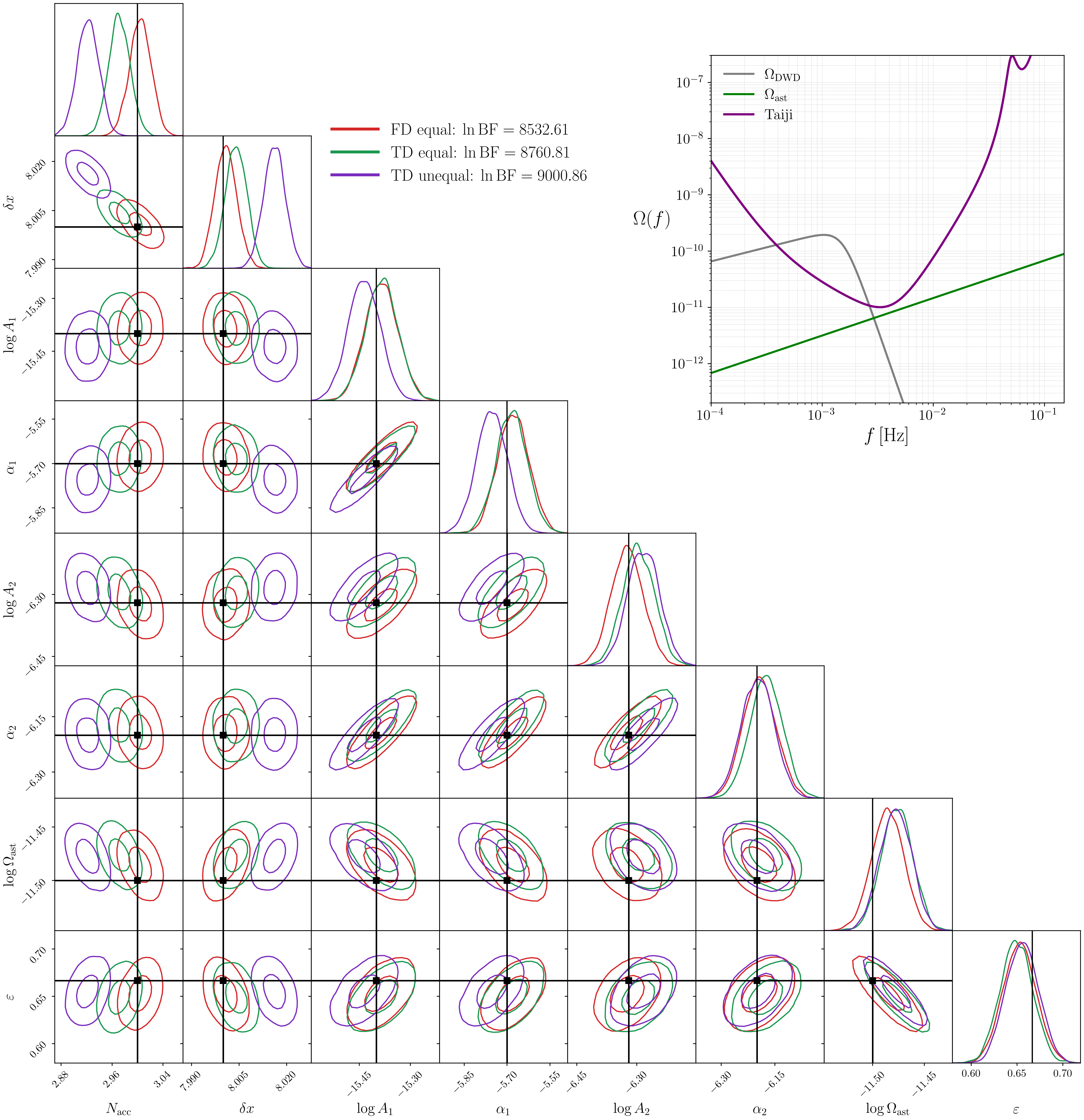}}
    \vspace{-0.75em}
    \caption{Corner plot for the Galactic foreground plus stochastic astrophysical background benchmark at $\log_{10}\Omega_{\rm ast}=-11.5$. Red, green, and purple contours denote the equal-arm FD, equal-arm TD, and unequal-arm TD posteriors, respectively. For each posterior, the inner and outer contours indicate the $1\sigma$ and $2\sigma$ credible regions, respectively. Black crosshairs mark the injected parameter values. The plotted parameter vector is $(N_{\rm acc},\delta x,\log_{10}A_1,\alpha_1,\log_{10}A_2,\alpha_2,\log_{10}\Omega_{\rm ast},\varepsilon)$. The upper-right inset shows the injected DWD foreground, stochastic astrophysical background, and effective noise curve; the annotations report the corresponding $\ln\mathrm{BF}$ values for the three PE configurations.}
    \label{fig:astro8corners}
\end{figure}

As a representative benchmark, we set $\log_{10}\Omega_{\rm ast}=-11.5$. Fig.~\ref{fig:astro8corners} compares the NS posterior samples for the equal-arm FD reference, the equal-arm TD analysis, and the unequal-arm TD analysis. The posterior distributions of $\log_{10}\Omega_{\rm ast}$ and $\varepsilon$ remain compact and centered near the injected values in all three configurations, even after the four DWD-foreground parameters are sampled jointly with the target component. Since the astrophysical background is modeled as a single power law, while the DWD foreground has a broken-power-law-like turnover shape, the two components can have partially similar spectral behavior and may therefore interfere with each other in the joint fit. Nevertheless, the posterior results show that this degeneracy is not strong enough to prevent recovery of the astrophysical-background parameters: both $\log_{10}\Omega_{\rm ast}$ and $\varepsilon$ are recovered consistently in all three configurations.

The largest visible shifts occur in the instrumental-noise parameters. This behavior is consistent with the frequency ranges in which the two noise contributions are most informative. The ACC contribution dominates the low-frequency part of the band, where finite-window leakage is more important and the frequency cut leaves fewer retained bins. The OMS contribution is mainly constrained at higher frequencies, but much of the rapidly oscillating structure near the TDI nulls is removed by the notch mask.

The detailed recovery results for the benchmark in Fig.~\ref{fig:astro8corners} are listed in Table~\ref{tab:astro8corner}. The FIM column gives the local relative uncertainty at the injected point, while the three posterior columns give the NS posterior means and relative posterior uncertainties. The injected value $\log_{10}\Omega_{\rm ast}=-11.5$ is recovered as $-11.485$, $-11.477$, and $-11.478$ in the equal-arm FD, equal-arm TD, and unequal-arm TD analyses, respectively. The recovered spectral indices also remain close to the injected value $\varepsilon=2/3$. Within the matched full-covariance model, this benchmark therefore shows that the astrophysical background remains identifiable after marginalizing over the effective Galactic foreground. The posterior means of the noise parameters show larger realization-dependent offsets, but these shifts do not propagate into a significant bias in the recovered astrophysical-background parameters. 

\FloatBarrier

\begin{table}[t!]
    \centering
    \tiny
    \setlength{\tabcolsep}{2.2pt}
    \renewcommand{\arraystretch}{0.96}
    \resizebox{\linewidth}{!}{
    \begin{tabular}{lcccccccc}
    \toprule
    Parameter &
    Injected &
    \makecell{FIM\\unc. (\%)} &
    \makecell{Equal-arm\\FD mean} &
    \makecell{Equal-arm\\FD unc. (\%)} &
    \makecell{Equal-arm\\TD mean} &
    \makecell{Equal-arm\\TD unc. (\%)} &
    \makecell{Unequal-arm\\TD mean} &
    \makecell{Unequal-arm\\TD unc. (\%)} \\
    \midrule
    $N_{\rm acc}/10^{-15}$ & $3.000$ & $0.565$ & $3.005$ & $0.568$ & $2.972$ & $0.554$ & $2.922$ & $0.563$ \\
    $\delta x/10^{-12}$ & $8.000$ & $0.043$ & $8.001$ & $0.043$ & $8.004$ & $0.042$ & $8.016$ & $0.043$ \\
    $\log_{10}A_1$ & $-15.400$ & $0.304$ & $-15.385$ & $0.312$ & $-15.384$ & $0.310$ & $-15.436$ & $0.303$ \\
    $\alpha_1$ & $-5.700$ & $0.931$ & $-5.678$ & $0.955$ & $-5.684$ & $0.945$ & $-5.753$ & $0.929$ \\
    $\log_{10}A_2$ & $-6.320$ & $0.620$ & $-6.324$ & $0.631$ & $-6.295$ & $0.622$ & $-6.283$ & $0.620$ \\
    $\alpha_2$ & $-6.200$ & $0.726$ & $-6.193$ & $0.740$ & $-6.174$ & $0.734$ & $-6.198$ & $0.730$ \\
    $\log_{10}\Omega_{\rm ast}$ & $-11.500$ & $0.134$ & $-11.485$ & $0.134$ & $-11.477$ & $0.131$ & $-11.478$ & $0.130$ \\
    $\varepsilon$ & $0.667$ & $2.705$ & $0.653$ & $2.706$ & $0.649$ & $2.635$ & $0.656$ & $2.617$ \\
    \bottomrule
    \end{tabular}
    }
    \caption{Parameter recovery for the Galactic foreground plus stochastic astrophysical background benchmark shown in Fig.~\ref{fig:astro8corners}. ``FIM unc.'' is computed from the inverse FIM. The equal-arm FD, equal-arm TD, and unequal-arm TD columns report NS posterior means and relative posterior uncertainties.}
    \label{tab:astro8corner}
\end{table}

\begin{figure}[t!]
    \centering
    \includegraphics[width=0.92\linewidth]{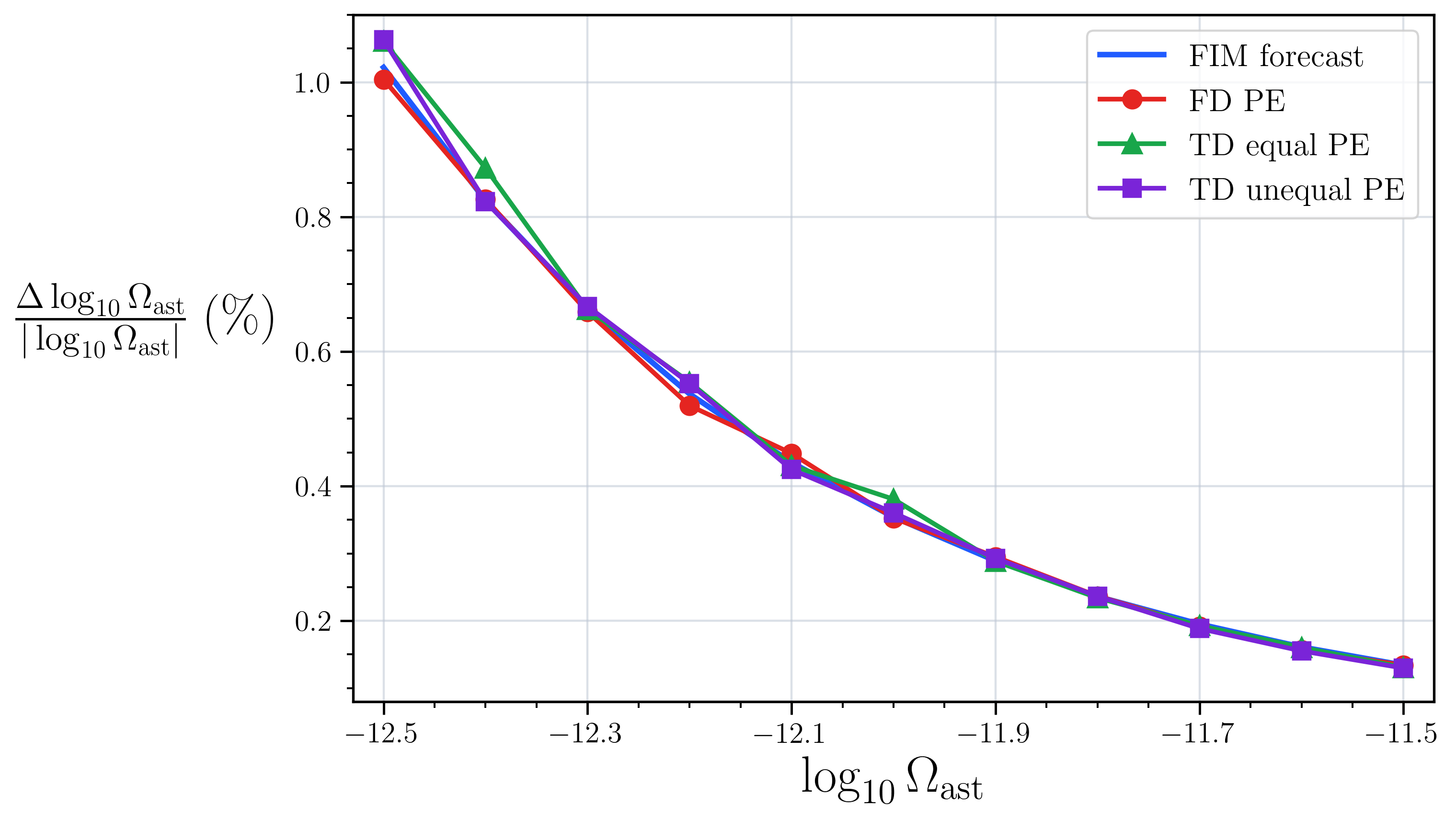}
    \caption{Relative uncertainty of $\log_{10}\Omega_{\rm ast}$ in the Galactic-foreground plus stochastic-astrophysical-background analysis as a function of the injected astrophysical-background amplitude. The blue curve shows the FIM prediction for the equal-arm FD configuration. Red circles, green triangles, and purple squares show the equal-arm FD, equal-arm TD, and unequal-arm TD posterior uncertainty estimates, respectively.}
    \label{fig:astro8unc}
\end{figure}

To quantify how the measurement precision changes as the astrophysical background weakens, we scan an observationally motivated compact-binary background interval. The component $\Omega_{{\rm GW},{\rm ast}}$ represents the unresolved extragalactic compact-binary background at the reference frequency $10^{-3}\,\mathrm{Hz}$, modeled with the same power-law form used in our previous analysis~\cite{Liang:2025zku,Guan:2025idx,Babak:2023lro}. Following Ref.~\cite{Liang:2025zku}, we use $\log_{10}\Omega_{\rm ast}=-11.5$ as the fiducial astrophysical-background amplitude. We take $\log_{10}\Omega_{\rm ast}=-12.5$ as a conservative weak-background endpoint, motivated by the compact-binary SGWB estimates discussed in Ref.~\cite{Boileau:2020rpg}. The scan is therefore restricted to
\begin{equation}
    -12.5 \le \log_{10}\Omega_{\rm ast} \le -11.5 .
\end{equation}

For each injected amplitude, we compare the FIM uncertainty predicted for the equal-arm FD reference with the posterior uncertainties obtained from NS in the three different configurations. Fig.~\ref{fig:astro8unc} shows the relative uncertainty of $\log_{10}\Omega_{\rm ast}$ as a function of the injected amplitude. Both the local FIM estimate and the posterior widths show the expected degradation: as the astrophysical background becomes weaker, the recovery becomes less precise and the relative uncertainty increases.

\begin{table}[t]
\centering
\fontsize{6.5pt}{7.5pt}\selectfont
\setlength{\tabcolsep}{4.0pt}
\renewcommand{\arraystretch}{1.2}
\resizebox{\linewidth}{!}{
\begin{tabular}{c cc ccc}
\toprule
$\log_{10}\Omega_{\rm ast}$ &
$\mathrm{SNR}_{\rm abs}$ &
$\mathrm{SNR}_{\rm rel}$ &
\makecell{Equal-arm FD\\$\ln\mathrm{BF}$} &
\makecell{Equal-arm TD\\$\ln\mathrm{BF}$} &
\makecell{Unequal-arm TD\\$\ln\mathrm{BF}$} \\
\midrule
$-11.5$ & $377$ & $332$ & $8532.61$ & $8760.81$ & $9000.86$ \\
$-11.6$ & $299$ & $263$ & $5838.71$ & $5873.19$ & $5857.24$ \\
$-11.7$ & $238$ & $209$ & $3953.07$ & $3920.16$ & $3964.49$ \\
$-11.8$ & $189$ & $166$ & $2597.61$ & $2813.89$ & $2502.61$ \\
$-11.9$ & $151$ & $132$ & $1690.29$ & $1795.41$ & $1624.62$ \\
$-12.0$ & $119$ & $105$ & $1131.98$ & $1004.69$ & $1131.04$ \\
$-12.1$ & $95$ & $83$ & $727.62$ & $735.43$ & $713.57$ \\
$-12.2$ & $76$ & $66$ & $495.25$ & $429.97$ & $493.70$ \\
$-12.3$ & $59$ & $52$ & $321.56$ & $308.78$ & $311.38$ \\
$-12.4$ & $47$ & $42$ & $183.12$ & $191.55$ & $185.62$ \\
$-12.5$ & $37$ & $33$ & $127.50$ & $115.76$ & $127.21$ \\
\bottomrule
\end{tabular}
}
\caption{Summary of SNRs and BFs for the Galactic-foreground plus stochastic-astrophysical-background scan over $-12.5 \le \log_{10}\Omega_{\rm ast} \le -11.5$. The two SNR columns are equal-arm local detectability diagnostics in the full $XYZ$ basis defined in Eq.~\eqref{eq:snrdefs}: $\mathrm{SNR}_{\rm abs}$ only uses the instrumental noise as the baseline, while $\mathrm{SNR}_{\rm rel}$ uses instrumental noise plus the unresolved Galactic foreground. The BFs are obtained from the corresponding \pkg{Bilby.dynesty} evidence outputs for the equal-arm FD analysis, the equal-arm TD simulated-data analysis, and the unequal-arm TD simulated-data analysis. Each entry compares the null and alternative models within the same analysis configuration.}
\label{tab:astro8summary}
\end{table}

The uncertainty degradation in Fig.~\ref{fig:astro8unc} is accompanied by the expected decline in SNR and evidence. Table~\ref{tab:astro8summary} summarizes the corresponding SNR and evidence diagnostics for the amplitude scan. The SNRs are computed with the equal-arm detector functions, whereas the BFs are listed separately for the three PE configurations. At the weakest injection, $\log_{10}\Omega_{\rm ast}=-12.5$, the absolute and relative SNRs are $37$ and $33$, respectively, and the corresponding values are $\ln\mathrm{BF}=127.50$, $115.76$, and $127.21$ for the equal-arm FD, equal-arm TD, and unequal-arm TD analyses.

\begin{figure}[t!]
    \centering
    \includegraphics[width=0.98\linewidth]{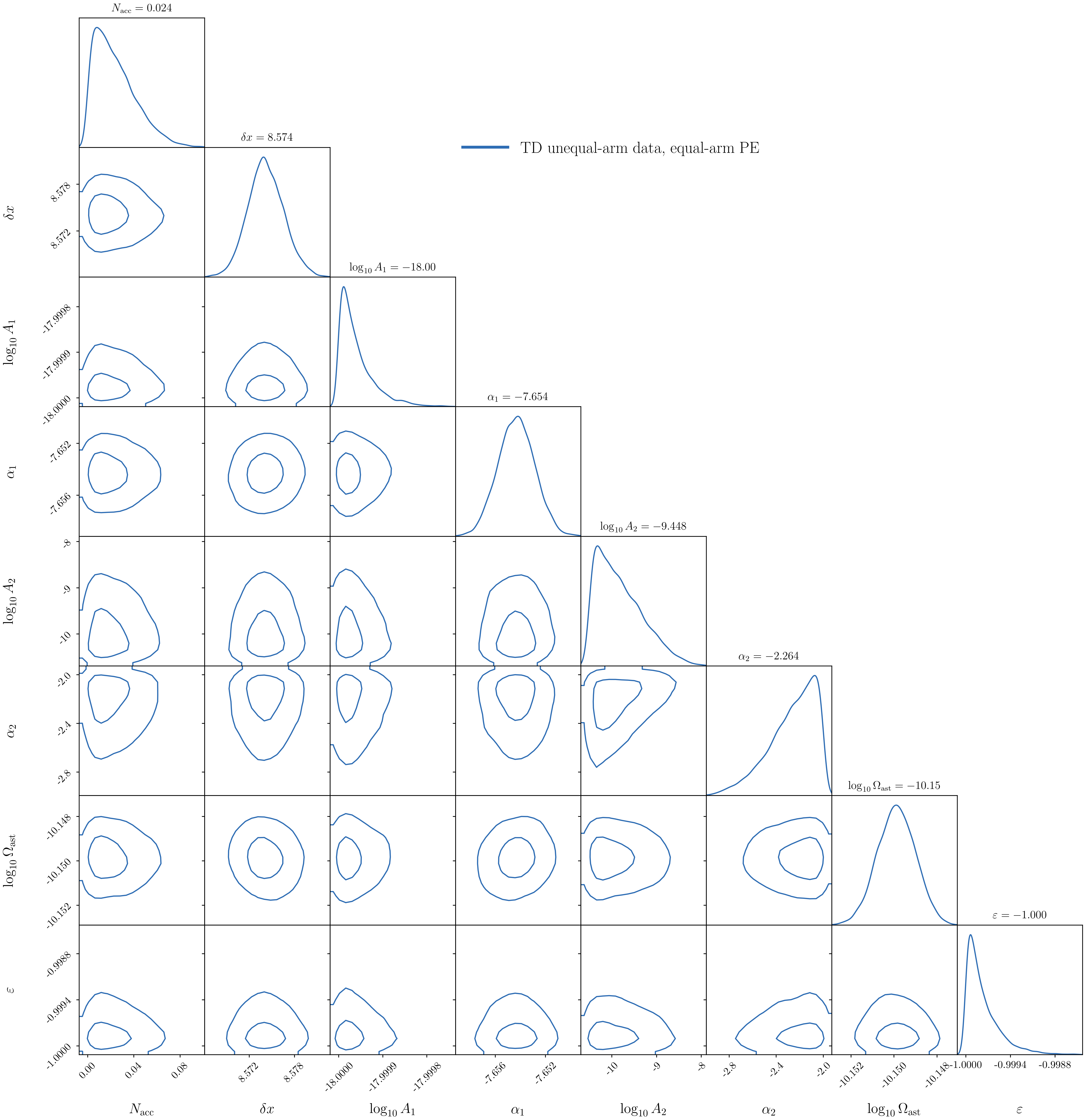}
    \caption{Deliberate model-mismatch PE example for the astrophysical-background benchmark. To assess the effect of detector-model mismatch, we generate the data using the unequal-arm TD configuration and analyze them with an equal-arm spectral model. The recovered values are listed above the one-dimensional posterior panels. The displaced and boundary-dominated posteriors illustrate why realistic TD data should be analyzed with segment-dependent response and noise transfer functions. For each posterior, the inner and outer contours indicate the $1\sigma$ and $2\sigma$ credible regions, respectively.}
    \label{fig:astro8mismatch}
\end{figure}

Taken together, the matched-pipeline comparison shows that the equal-arm FD framework provides a useful simplified baseline for our analysis. The explicit TD equal-arm and TD unequal-arm pipelines include segmentation, windowing, masking, and segment-wise response and noise transfer modeling, yet they yield comparable posterior precision, SNRs, and BF diagnostics for the astrophysical-background amplitude over the amplitude range considered here. Because the TD unequal-arm pipeline is closest to the realistic detector configuration, its agreement with the FD reference indicates that the equal-arm FD simplification is adequate for the signal-recovery and detectability conclusions analyzed here. The remaining differences among configurations, especially at the weakest amplitudes, are consistent with finite-realization and sampler scatter and do not show a systematic FD/TD or equal-arm/unequal-arm bias.

\FloatBarrier

To assess the consequences of a detector-model mismatch, we analyze data generated with the unequal-arm TD configuration using an equal-arm spectral model. This test uses the same injected TD unequal-arm data set as the representative benchmark above, with the same injected parameters, prior ranges, and other analysis settings. The only change is that the data are now analyzed with a mismatched equal-arm PE model instead of the corresponding unequal-arm and segment-dependent model. Therefore, Fig.~\ref{fig:astro8mismatch} is a controlled model-mismatch comparison designed to isolate the effect of using an inconsistent inference model. As a result, several instrumental and foreground parameters become significantly displaced or boundary dominated.

\begin{table}[t!]
    \centering
    \tiny
    \setlength{\tabcolsep}{2.2pt}
    \renewcommand{\arraystretch}{0.96}
    \resizebox{\linewidth}{!}{
    \begin{tabular}{lcccccccc}
    \toprule
    Parameter & Injected & \makecell{FIM\\unc. (\%)} & \makecell{FD eq.\\mean} & \makecell{FD eq.\\unc. (\%)} & \makecell{TD eq.\\mean} & \makecell{TD eq.\\unc. (\%)} & \makecell{TD uneq.\\mean} & \makecell{TD uneq.\\unc. (\%)} \\
    \midrule
    $N_{\rm acc}/10^{-15}$ & $3.000$ & $0.587$ & $2.991$ & $0.587$ & $3.021$ & $0.592$ & $2.956$ & $0.579$ \\
    $\delta x/10^{-12}$ & $8.000$ & $0.044$ & $7.998$ & $0.043$ & $7.996$ & $0.044$ & $8.007$ & $0.043$ \\
    $\log_{10}A_1$ & $-15.400$ & $0.591$ & $-15.384$ & $0.563$ & $-15.404$ & $0.597$ & $-15.487$ & $0.601$ \\
    $\alpha_1$ & $-5.700$ & $1.686$ & $-5.688$ & $1.616$ & $-5.709$ & $1.708$ & $-5.786$ & $1.710$ \\
    $\log_{10}A_2$ & $-6.320$ & $1.161$ & $-6.276$ & $1.105$ & $-6.296$ & $1.170$ & $-6.404$ & $1.188$ \\
    $\alpha_2$ & $-6.200$ & $1.337$ & $-6.170$ & $1.279$ & $-6.188$ & $1.352$ & $-6.281$ & $1.364$ \\
    $\log_{10}\Omega_{\rm ast}$ & $-11.500$ & $0.870$ & $-11.543$ & $0.861$ & $-11.495$ & $0.842$ & $-11.472$ & $0.852$ \\
    $\varepsilon$ & $0.667$ & $13.742$ & $0.715$ & $13.501$ & $0.661$ & $13.286$ & $0.650$ & $13.328$ \\
    $\log_{10}\Omega_0$ & $-11.500$ & $0.918$ & $-11.487$ & $0.899$ & $-11.531$ & $0.958$ & $-11.539$ & $1.137$ \\
    $\log_{10}(f_p/\mathrm{Hz})$ & $-2.250$ & $1.222$ & $-2.257$ & $1.155$ & $-2.246$ & $1.390$ & $-2.285$ & $1.839$ \\
    \bottomrule
    \end{tabular}
    }
    \caption{Parameter recovery for the ten-parameter SW benchmark shown in Fig.~\ref{fig:sw10corners}. The FIM forecast and all posterior estimates use the same full covariance model in the $XYZ$ basis. The equal-arm FD, equal-arm TD, and unequal-arm TD columns report NS posterior means and relative posterior uncertainties for the $\log_{10}\Omega_0=-11.5$ benchmark.}
    \label{tab:sw10corner}
\end{table}

In this mismatch run, the astrophysical-background amplitude is inferred under a covariance model that is inconsistent with the TD unequal-arm data. The result confirms the need for internal consistency between simulation and inference: data generated with time-dependent unequal arms should be analyzed with segment-dependent unequal-arm response and noise transfer functions.

\subsection{Phase-Transition Gravitational-Wave Background Search}\label{sec:swsrch}

We now include the SW component in Eq.~\eqref{eq:swmodel}. The full model contains instrumental noise, the unresolved Galactic foreground, the stochastic astrophysical background, and the SW spectrum \cite{Huang:2025uer,Boileau:2022ter}. The null model used in the BF comparison contains the same components, but with the SW contribution removed. The sampled parameter vector is
\begin{equation}
\theta_{\rm SW}=\big(N_{\rm acc},\delta x,\log_{10}A_1,\alpha_1,\log_{10}A_2,\alpha_2, \log_{10}\Omega_{\rm ast},\varepsilon,\log_{10}\Omega_0,\log_{10}(f_p/\mathrm{Hz})\big).
\end{equation}

\begin{figure}[t!]
    \centering
    \makebox[\linewidth][c]{\includegraphics[width=1.18\linewidth]{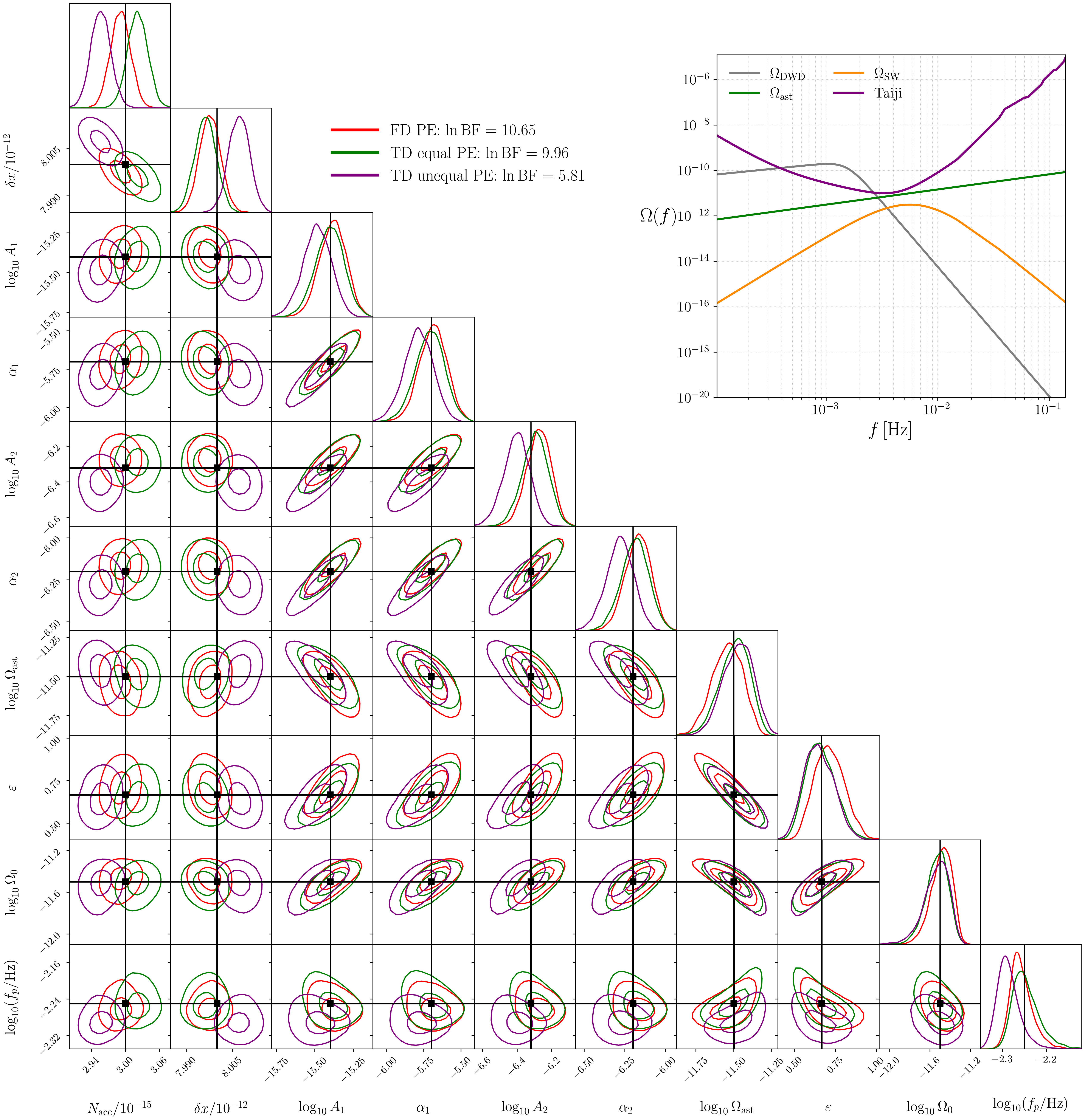}}
    \caption{Ten-parameter SW benchmark corner plot for $\log_{10}\Omega_0=-11.5$ and $\log_{10}(f_p/\mathrm{Hz})=-2.25$. Red, green, and purple contours denote the equal-arm FD, equal-arm TD, and unequal-arm TD posteriors, respectively. For each posterior, the inner and outer contours indicate the $1\sigma$ and $2\sigma$ credible regions, respectively. Black crosshairs mark the injected parameter values. The plotted parameter vector is $\theta_{\rm SW}$, as defined in the text. The upper-right inset shows the DWD foreground, stochastic astrophysical background, SW spectrum, and effective noise curve used for this benchmark. The BF annotations give the evidence ratios for the three PE configurations: $\ln{\rm BF}=10.65$, $9.96$, and $5.81$.}
    \label{fig:sw10corners}
\end{figure}

\FloatBarrier

This ten-parameter model extends the analysis of Sec.~\ref{sec:astrofgbg}. The two additional SW parameters make the inference more challenging because the peak amplitude and peak frequency can partially trade off against both the amplitude and slope of the astrophysical background, as well as against the turnover structure of the effective Galactic foreground.

As a representative benchmark, we set $\log_{10}\Omega_0=-11.5$ and $\log_{10}(f_p/\mathrm{Hz})=-2.25$, with all remaining fiducial values listed in Table~\ref{tab:pepriors}. The parameter-level recovery for this benchmark is summarized in Table~\ref{tab:sw10corner}.

Fig.~\ref{fig:sw10corners} compares the posterior samples from the equal-arm FD, equal-arm TD, and unequal-arm TD analyses for the ten-parameter SW benchmark. Table~\ref{tab:sw10corner} reports the corresponding numerical recovery values: the FIM column provides a local equal-arm FD reference, while the NS posterior columns show how the TD treatment and unequal-arm effects shift the recovered means and relative uncertainties.

\section{Conclusions}\label{sec:conclusion}

We have implemented a full-covariance Bayesian spectral-inference framework for SGWB studies with Taiji-like missions. Second-generation Michelson \(X,Y,Z\) streams are generated in TD, divided into finite segments, Fourier transformed, and analyzed with a FD likelihood that retains the segment-dependent complex \(3\times3\) spectral covariance. This construction combines realistic TDI data generation with Bayesian inference while incorporating the orbit-dependent response functions and ACC and OMS noise transfer functions. In controlled component-wise simulations, the calculated detector functions reproduce the corresponding auto- and cross-spectra within the finite-realization scatter expected for the averaged spectral estimators over the retained band away from TDI nulls. The four-parameter recovery tests, repeated over ten independent TD realizations, further show that the same full-covariance detector model yields approximately unbiased parameter estimates.

The eight-parameter astrophysical-background analysis tests the framework in a more crowded spectral environment. After simultaneously marginalizing over the instrumental-noise amplitudes and the four parameters of the effective Galactic DWD foreground, both the amplitude and spectral index of the stochastic astrophysical background remain identifiable. Over the interval \(-12.5\leq\log_{10}\Omega_{\rm ast}\leq-11.5\), the static equal-arm FD, equal-arm TD, and unequal-arm TD configurations yield comparable posterior precision and consistent SNR and BF trends. The precision and evidence decrease as the injected background weakens, but the astrophysical component remains strongly favored throughout the range considered. The agreement among the matched configurations establishes the equal-arm FD treatment as a useful simplified baseline for this benchmark. TD simulations are nevertheless essential because they incorporate the TD unequal-arm geometry and therefore provide a more realistic description of the detector. In addition, the deliberate mismatch test produces strongly displaced posteriors when unequal-arm TD data are analyzed with a static equal-arm covariance, demonstrating the importance of maintaining consistency between the simulated data and the inference model.

Finally, extending the analysis to include an SW spectrum produces a ten-parameter benchmark containing instrumental noise, the Galactic foreground, the astrophysical background, and a cosmological component. The SW peak amplitude and frequency are recovered near their injected values in all three matched configurations. Future work should incorporate time-dependent and nonstationary instrumental noise, data gaps, calibration uncertainties, and more realistic Galactic-foreground models \cite{Buscicchio:2025zeb,Pozzoli:2025hhl,Alvey:2024uoc}. Embedding this detector-level treatment in a global-fit analysis would ultimately allow stochastic backgrounds, instrumental noise, Galactic binaries, and individually resolvable sources to be inferred within a common framework.

\FloatBarrier

\begin{appendices}
\setcounter{figure}{0}
\renewcommand{\thefigure}{\thesection.\arabic{figure}}

\section{Detector-Function Validation Plots}\label{app:detector-validation}

This appendix collects the detailed validation plots for the detector response and noise transfer functions summarized in Sec.~\ref{sec:tdsimvalidation}. The simulated points are formed from the single-realization estimator in Eq.~\eqref{eq:observedcsd}. Averaging over \(N_{\rm real}=1000\) independent realizations and removing the input normalization parameters gives
\begin{equation}
\begin{aligned}
    \widehat{\mathcal N}^{\rm acc,\kappa}_{IJ}(f_k)
    &=
    \frac{1}{N_{\rm real}N_{\rm acc}^2}
    \sum_{n=1}^{N_{\rm real}}
    \widehat G_{IJ,n}^{\kappa,{\rm acc}}(f_k),\\
    \widehat{\mathcal N}^{\rm OMS,\kappa}_{IJ}(f_k)
    &=
    \frac{1}{N_{\rm real}\delta x^2}
    \sum_{n=1}^{N_{\rm real}}
    \widehat G_{IJ,n}^{\kappa,{\rm OMS}}(f_k),\\
    \widehat{\mathcal R}^{\kappa}_{IJ}(f_k)
    &=
    \frac{1}{N_{\rm real}}
    \sum_{n=1}^{N_{\rm real}}
    \widehat G_{IJ,n}^{\kappa,{\rm resp}}(f_k).
\end{aligned}
\end{equation}
The superscript on \(\widehat G_{IJ,n}\) identifies the isolated \trianglesim{} run. The factors \(N_{\rm acc}^2\) and \(\delta x^2\) remove the single-link amplitude normalizations from the ACC and OMS spectra, respectively. No additional amplitude normalization is required for the response run because the injected strain PSD is set to the unit value, $S_h(f)=1\,\mathrm{Hz}^{-1}$, in the adopted one-sided convention.

\begin{figure}[t]
    \centering
    \makebox[\linewidth][c]{\includegraphics[width=1.18\linewidth]{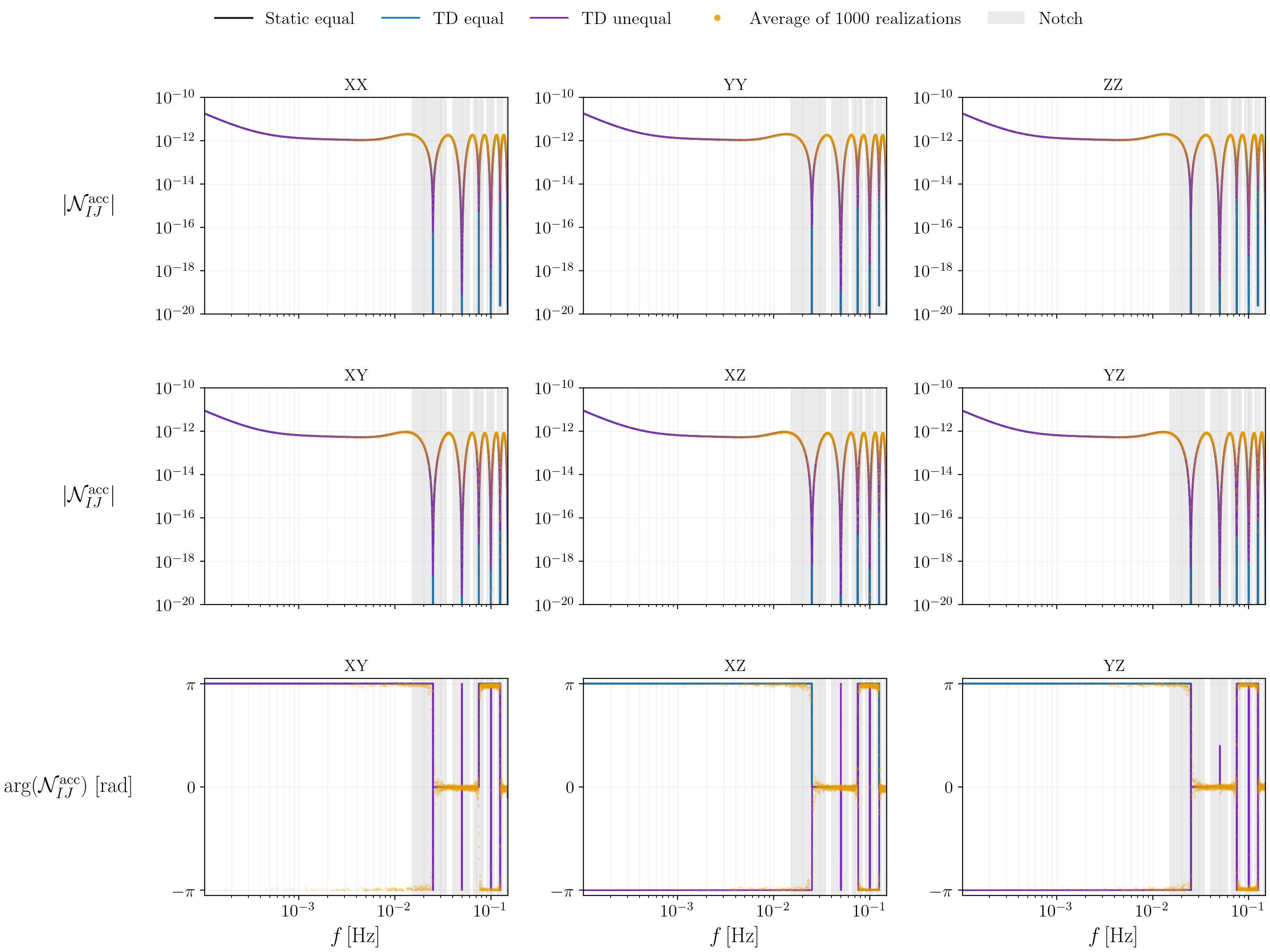}}
    \caption{
    ACC noise transfer function validation with the average of 1000 realizations overlaid. The plotted object is the ACC noise transfer function in the $XYZ$ basis, with the ACC noise amplitude factored out. The first row shows the amplitudes of the auto spectra $(XX,YY,ZZ)$, the middle row shows the amplitudes of the cross spectra $(XY,XZ,YZ)$, and the bottom row shows the phases of those cross spectra. The black curve gives the static equal-arm reference, while the blue and purple curves show the equal-arm TD and unequal-arm TD calculations with the same convention in the first segment. Curves that are not separately visible are visually coincident with another curve at the plotted resolution. The orange points are the average of 1000 \trianglesim{} TD realizations with only ACC noise for the first segment of the Taiji unequal-arm orbit. Pale gray bands mark the TDI notch neighborhoods, while the curves and points are still drawn through those frequencies to show where the transfer function zeros make the phase most rapidly varying. Overall, the amplitude panels agree very well, and the visible deviations are dominated by phase fluctuations, especially near the TDI null points.
    }
    \label{fig:accfdtd}
\end{figure}

\begin{figure}[t]
    \centering
    \makebox[\linewidth][c]{\includegraphics[width=1.18\linewidth]{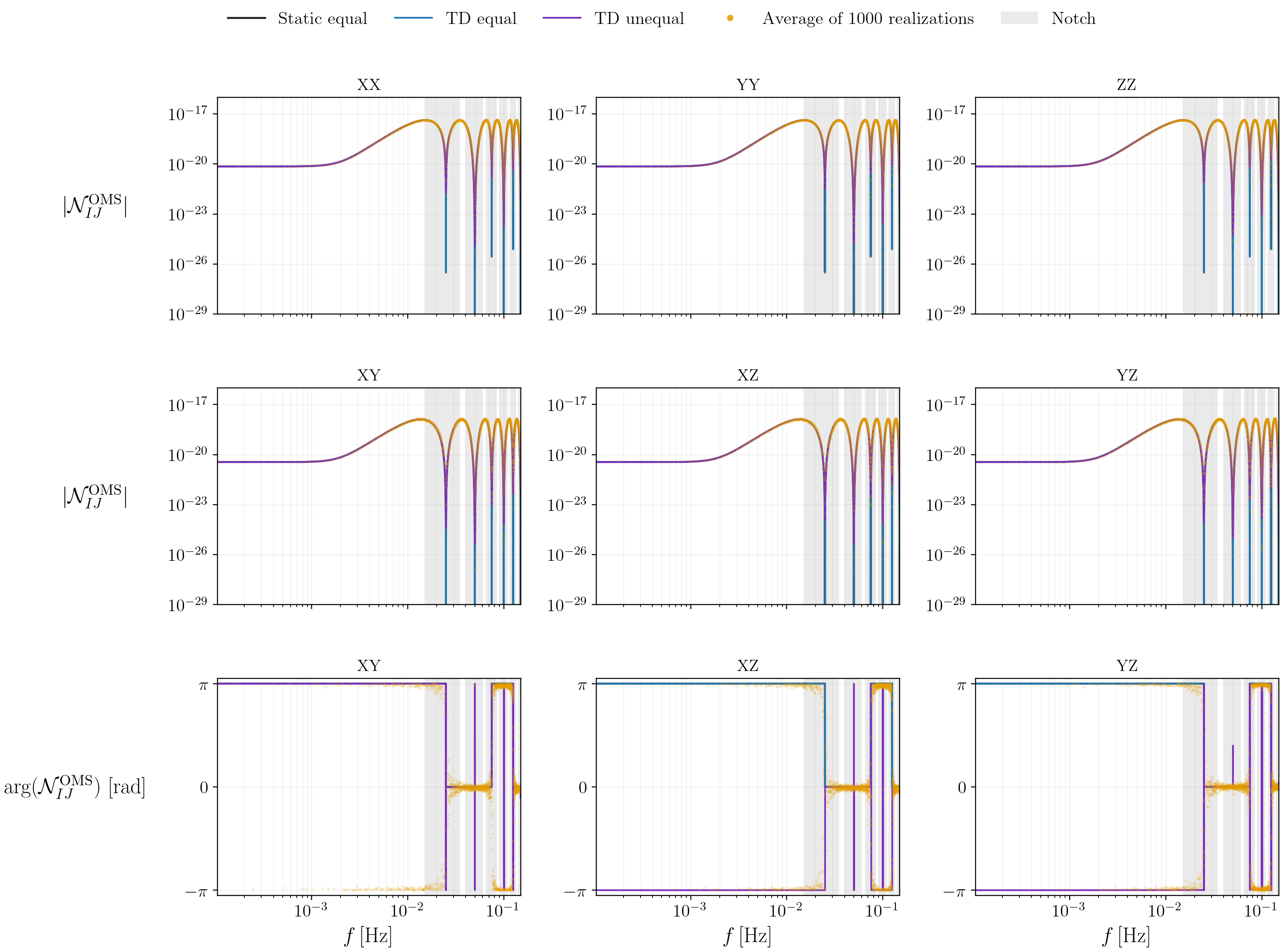}}
    \caption{
    OMS noise transfer function validation. The plotted object is the OMS noise transfer function in the $XYZ$ basis, with the OMS noise amplitude factored out. All plotting conventions, simulation settings, curve labels, orange realization averages, and pale-gray TDI notch markers are the same as in Fig.~\ref{fig:accfdtd}. As in the ACC case, the amplitude panels agree very well, and the visible deviations are dominated by phase fluctuations, especially near the TDI null points.
    }
    \label{fig:omsfdtd}
\end{figure}

\begin{figure}[t]
    \centering
    \makebox[\linewidth][c]{\includegraphics[width=1.12\linewidth]{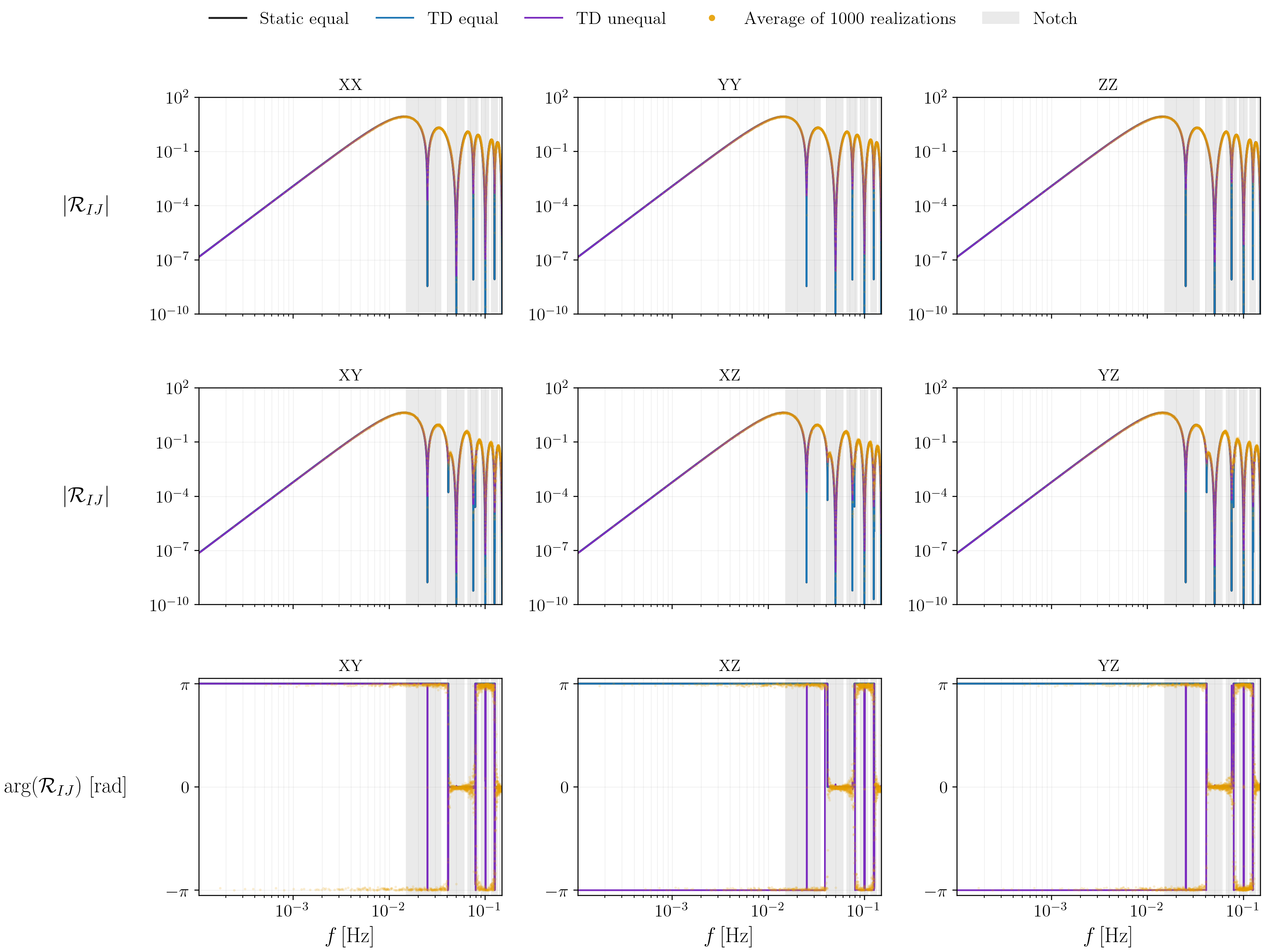}}
    \caption{
    Response function validation with controlled response-only simulation points overlaid. The plotted object is the full response function in the $XYZ$ basis that maps an input SGWB strain spectrum into the detector PSD matrix. The panel ordering, curve labels, orange realization averages, and pale-gray TDI notch markers follow the same conventions as Fig.~\ref{fig:accfdtd}. Away from the null neighborhoods the agreement remains good. Near the TDI null points, the response comparison shows stronger fluctuations than Figs.~\ref{fig:accfdtd} and~\ref{fig:omsfdtd}, with more visible differences in both phase and amplitude.
    }
    \label{fig:responsefdtd}
\end{figure}

Figs.~\ref{fig:accfdtd}--\ref{fig:responsefdtd} display the ACC noise transfer functions, the OMS noise transfer functions, and the response functions, respectively, using a common nine-panel \(XYZ\) layout. The panels are organized by channel pair: the first row shows the diagonal amplitudes \((XX,YY,ZZ)\), the second row shows the off-diagonal amplitudes \((XY,XZ,YZ)\), and the third row shows the off-diagonal phases. In these overlay plots, the black, blue, and purple curves denote the calculated static equal-arm FD, equal-arm TD, and unequal-arm TD functions, while the orange markers show the corresponding 1000-realization mean from the first-segment unequal-arm TD validation run. The static equal-arm FD and equal-arm TD curves are therefore reference configurations, whereas the direct simulation-to-calculation comparison is confined to the first-segment unequal-arm TD run.

\begin{figure}[t]
    \centering
    \makebox[\linewidth][c]{\includegraphics[width=1.18\linewidth]{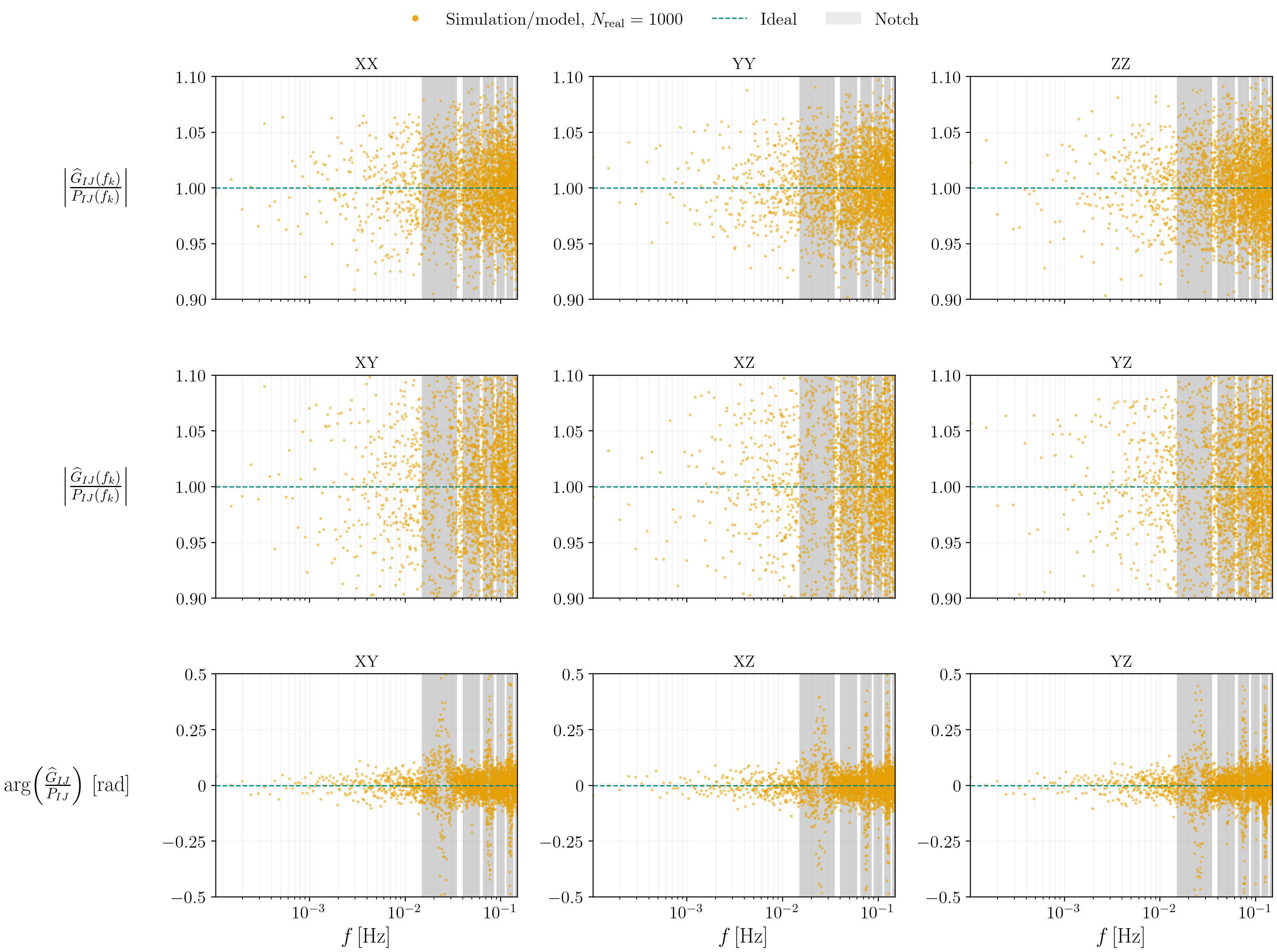}}
    \caption{
    Direct ratio view of the ACC noise transfer function validation in Fig.~\ref{fig:accfdtd}. This figure reuses the same data products shown there. In this caption, \(\widehat G\) denotes the averaged simulated estimate from the 1000 ACC-only TD realizations for the first segment after the injected ACC amplitude is divided out, while \(P\) denotes the corresponding calculated first-segment unequal-arm ACC noise transfer function on the same one-segment frequency grid. The first row shows \(|\widehat G_{II}/P_{II}|\) for the three diagonal entries, the second row shows \(|\widehat G_{IJ}/P_{IJ}|\) for the off-diagonal entries, and the third row shows \(\arg(\widehat G_{IJ}/P_{IJ})\) for the off-diagonal phase residuals. The green dashed line marks the ideal reference value, unity for the amplitude ratios and zero for the phase residuals. The pale-gray bands mark the TDI null neighborhoods, where the transfer functions approach their null points and finite-realization fluctuations become more visible in the ratios and phase residuals. Away from these null neighborhoods, the amplitude ratios remain close to unity, typically within \(0.95\)--\(1.05\), indicating consistent recovery of the ACC normalization and the off-diagonal cross terms. The visible departures are mainly confined to the null neighborhoods and nearby high-frequency null features.
    }
    \label{fig:accsimgrid}
\end{figure}

\begin{figure}[t]
    \centering
    \makebox[\linewidth][c]{\includegraphics[width=1.22\linewidth]{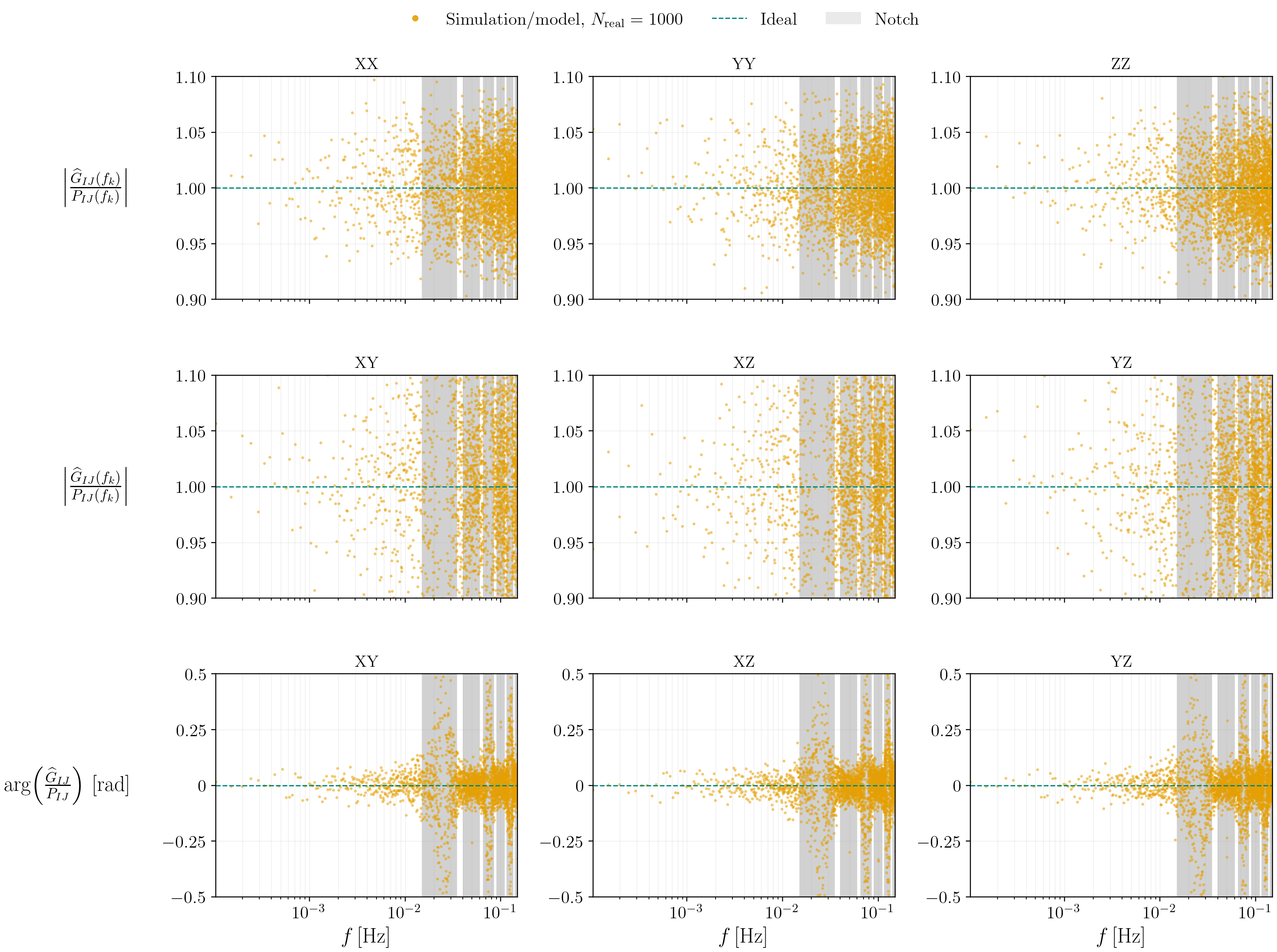}}
    \caption{
    Same ratio diagnostic and plotting convention as Fig.~\ref{fig:accsimgrid}, including the green dashed ideal reference line, now applied to the OMS noise transfer function validation in Fig.~\ref{fig:omsfdtd}. The numerator is the same mean of 1000 TD realizations with only OMS noise, and the denominator is the same calculated first-segment unequal-arm OMS transfer function on the same Fourier grid. The main cloud again stays close to unity, typically at the few percent level, while the most visible deviations are localized near the pale-gray TDI null neighborhoods.
    }
    \label{fig:omssimgrid}
\end{figure}

\begin{figure}[t]
    \centering
    \makebox[\linewidth][c]{\includegraphics[width=1.22\linewidth]{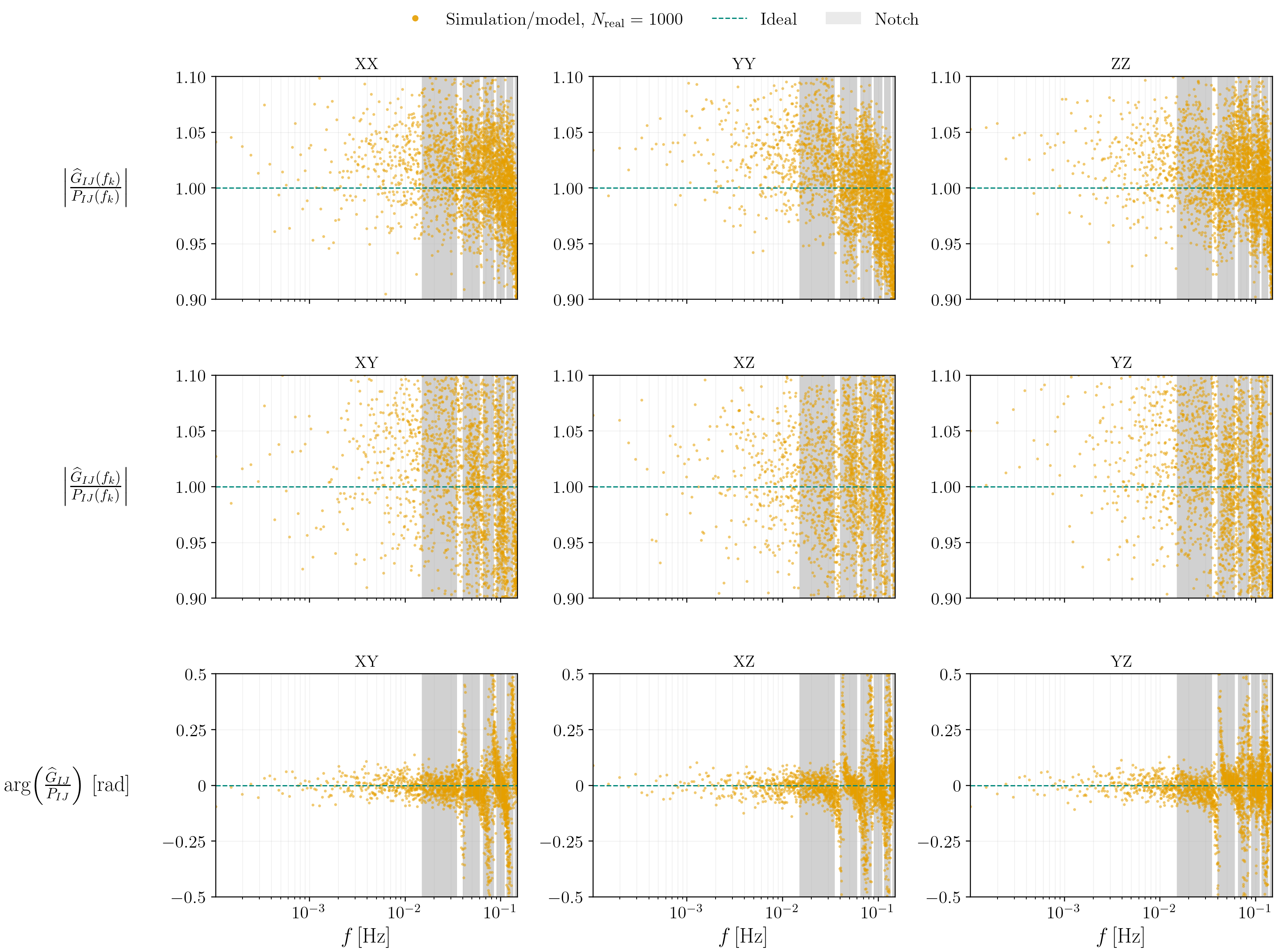}}
    \caption{
    Same ratio diagnostic and plotting convention as Fig.~\ref{fig:accsimgrid}, including the green dashed ideal reference line, now applied to the response function validation in Fig.~\ref{fig:responsefdtd}. The numerator is the same mean spectral response measured from 1000 controlled response-only TD realizations, and the denominator is the same calculated first-segment unequal-arm response function on the same Fourier grid. Most of the retained band remains close to the ideal value, with amplitude ratios typically within a few percent of unity. Compared with the ACC and OMS transfer functions, the ratios of response functions show stronger excursions near the TDI null neighborhoods, and the amplitude differences around those null points are more visible.
    }
    \label{fig:responsesimgrid}
\end{figure}

\begin{figure}[t]
    \centering
    \makebox[\linewidth][c]{\includegraphics[width=1.15\linewidth]{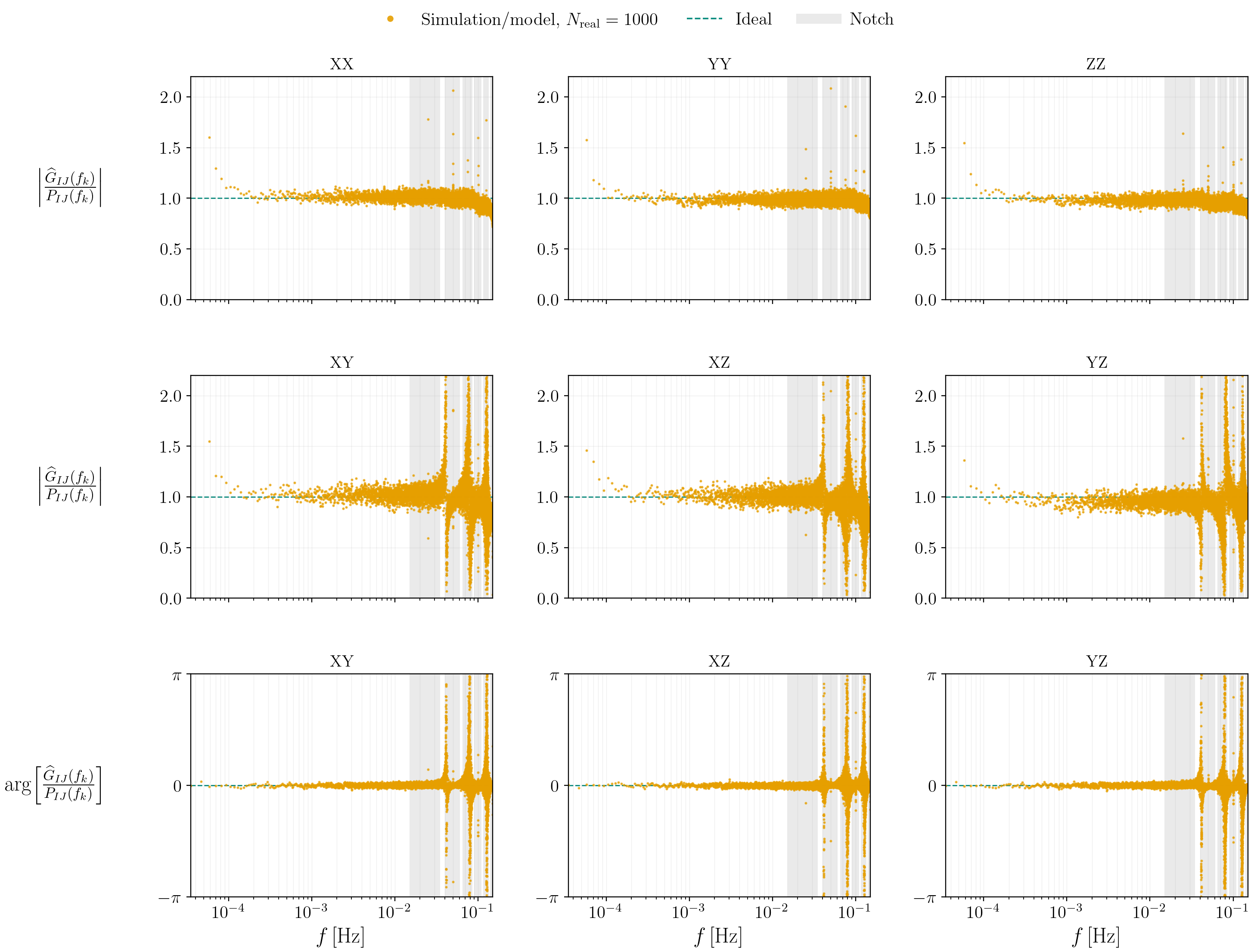}}
    \caption{
    First-segment spectral-function ratio in the full $XYZ$ basis for the 1000-realization unequal-arm TD simulation, with the stochastic astrophysical background, ACC noise, and OMS noise included simultaneously. Here, $\widehat G$ denotes the realization-averaged spectral estimate obtained from the simulated TD streams, while $P$ denotes the calculated first-segment auto-spectral density or CSD for the same $XYZ$ channel pair. The first row shows $|\widehat G_{II}(f_k)/P_{II}(f_k)|$ for the diagonal entries. The second row shows $|\widehat G_{IJ}(f_k)/P_{IJ}(f_k)|$ for the three independent off-diagonal amplitudes, and the third row shows $\arg[\widehat G_{IJ}(f_k)/P_{IJ}(f_k)]$ for the off-diagonal phases. The orange markers show the simulation-to-model ratios, or phase residuals, at each retained frequency bin. The green dashed line marks the ideal reference value, unity for amplitude ratios and zero for phase residuals; departures from this line show the residual scatter left after averaging. The pale-gray bands mark the $\pm0.010\,\mathrm{Hz}$ notch neighborhoods around the TDI null points. Near these null points, the transfer functions become small and rapidly varying, which amplifies finite-realization scatter residuals. The corresponding bins are therefore excluded by the fixed notch mask used in the inference analyses.
    }
    \label{fig:xyzcsdratio}
\end{figure}

Figs.~\ref{fig:accsimgrid}--\ref{fig:responsesimgrid} use the same ACC-only, OMS-only, and response-only validation averages to plot complex simulation-to-calculation ratios relative to the calculated first-segment unequal-arm TD functions. Fig.~\ref{fig:xyzcsdratio} applies the same ratio diagnostic to the full first-segment \(XYZ\) covariance, with ACC noise, OMS noise, and the stochastic astrophysical background included simultaneously.

\clearpage
\section{Cross-Spectral Residual Estimate}\label{app:rij-derivation}

This appendix gives the derivation of the reference level used for the squared normalized residual \(r_{IJ}^{\kappa}\) in Eq.~\eqref{eq:crossspectrum_relative_residual}. At a fixed segment \(\kappa\) and frequency bin \(f_k\), let \(n\) label one independent realization and write the normalized cross-spectrum estimator as
\begin{equation}
    \widehat G_{IJ,n}^{\kappa}(f_k)
    =
    \tilde d_{I,n}^{\kappa}(f_k)
    \tilde d_{J,n}^{\kappa}(f_k)^* ,
    \qquad
    \left\langle
    \widehat G_{IJ,n}^{\kappa}(f_k)
    \right\rangle
    =
    P_{IJ}^{\kappa}(f_k),
\end{equation}
where the same one-sided Fourier normalization as in Eq.~\eqref{eq:observedcsd} has already been applied. We suppress the common \((\kappa,f_k)\) arguments for simplification and write
\begin{equation}
    \widehat G_{IJ,n}=d_{I,n}d_{J,n}^*,
    \qquad
    P_{IJ}=\left\langle d_Id_J^*\right\rangle .
\end{equation}
For the complex Gaussian Fourier coefficients used here,
\begin{equation}
    \left\langle d_I d_J^* \right\rangle=P_{IJ},
    \qquad
    \left\langle d_I^* d_J \right\rangle=P_{JI},
    \qquad
    \left\langle d_I d_J \right\rangle
    =
    \left\langle d_I^* d_J^* \right\rangle=0,
\end{equation}
with $P_{JI}=P_{IJ}^*$. The cross-spectral covariance follows from the complex fourth moment, equivalently the complex Wishart covariance~\cite{Goodman:1963complex}. With the CSD and PSD convention used in the SGWB cross-correlation literature~\cite{Allen:1997ad,Romano:2016dpx}, the single-realization complex fluctuation is
\begin{equation}
    \left\langle
    \left|\widehat G_{IJ,n}-P_{IJ}\right|^2
    \right\rangle
    =
    \left\langle
    (d_Id_J^*-P_{IJ})(d_I^*d_J-P_{IJ}^*)
    \right\rangle  
    =
    \left\langle d_Id_J^*d_I^*d_J\right\rangle
    -|P_{IJ}|^2 .
\end{equation}
The remaining fourth moment is reduced by Wick's theorem for complex Gaussian variables. The three pairings give
\begin{equation}
\begin{aligned}
    \left\langle d_Id_J^*d_I^*d_J\right\rangle
    &=
    \left\langle d_Id_J^*\right\rangle
    \left\langle d_I^*d_J\right\rangle
    +
    \left\langle d_Id_I^*\right\rangle
    \left\langle d_J^*d_J\right\rangle
    +
    \left\langle d_Id_J\right\rangle
    \left\langle d_J^*d_I^*\right\rangle  \\
    &=
    P_{IJ}P_{JI}+P_{II}P_{JJ}+0
    =
    |P_{IJ}|^2+P_{II}P_{JJ}.
\end{aligned}
\end{equation}
The \(|P_{IJ}|^2\) term cancels the subtraction above, leaving
\begin{equation}\label{tgm}
    \left\langle
    \left|
    \widehat G_{IJ,n}^{\kappa}(f_k)
    -
    P_{IJ}^{\kappa}(f_k)
    \right|^2
    \right\rangle
    =
    P_{II}^{\kappa}(f_k)P_{JJ}^{\kappa}(f_k).
\end{equation}
Thus the absolute complex fluctuation of a single cross-spectrum estimate is controlled by the two auto spectra \(P_{II}^{\kappa}\) and \(P_{JJ}^{\kappa}\). It is not controlled by the magnitude of the mean cross spectrum \(P_{IJ}^{\kappa}\) alone. This point is important for off-diagonal elements, whose mean can be small even when the two auto spectra are not.

Following the validation setup described in Sec.~\ref{sec:tdsimvalidation}, the finite-realization mean used in the validation plots is
\begin{equation}
    \widehat G_{IJ}^{\kappa}(f_k)
    =
    \frac{1}{N_{\rm real}}
    \sum_{n=1}^{N_{\rm real}}
    \widehat G_{IJ,n}^{\kappa}(f_k),
\end{equation}
where \(N_{\rm real}\) is the number of independent realizations included in the average. Define the zero-mean fluctuation
\begin{equation}\label{wkx}
    \delta G_{IJ,n}^{\kappa}(f_k)
    =
    \widehat G_{IJ,n}^{\kappa}(f_k)
    -
    P_{IJ}^{\kappa}(f_k),
    \qquad
    \left\langle \delta G_{IJ,n}^{\kappa}(f_k)\right\rangle=0 .
\end{equation}
Then
\begin{equation}\label{whb}
    \widehat G_{IJ}^{\kappa}(f_k)-P_{IJ}^{\kappa}(f_k)
    =
    \frac{1}{N_{\rm real}}
    \sum_{n=1}^{N_{\rm real}}
    \delta G_{IJ,n}^{\kappa}(f_k).
\end{equation}
The squared fluctuation of the average is therefore
\begin{equation}\label{uyu}
\begin{aligned}
    \left\langle
    \left|\widehat G_{IJ}^{\kappa}(f_k)-P_{IJ}^{\kappa}(f_k)\right|^2
    \right\rangle
    &=
    \frac{1}{N_{\rm real}^2}
    \sum_{n,m}
    \left\langle
    \delta G_{IJ,n}^{\kappa}(f_k)
    \delta G_{IJ,m}^{\kappa}(f_k)^*
    \right\rangle .
\end{aligned}
\end{equation}
For $n\ne m$, different realizations are statistically independent; since each fluctuation has zero mean,
\begin{equation}\label{hbh}
    \left\langle
    \delta G_{IJ,n}^{\kappa}(f_k)
    \delta G_{IJ,m}^{\kappa}(f_k)^*
    \right\rangle
    =
    \left\langle
    \delta G_{IJ,n}^{\kappa}(f_k)
    \right\rangle
    \left\langle
    \delta G_{IJ,m}^{\kappa}(f_k)^*
    \right\rangle
    =
    0 .
\end{equation}
For a fixed channel pair $IJ$, segment $\kappa$, and frequency bin $f_k$, substituting Eqs.~\eqref{whb} and \eqref{hbh} into Eq.~\eqref{uyu} yields
\begin{equation}\label{slx}
\begin{aligned}
    \left\langle
    \left|\widehat G_{IJ}^{\kappa}(f_k)-P_{IJ}^{\kappa}(f_k)\right|^2
    \right\rangle
    &=
    \frac{1}{N_{\rm real}^2}
    \sum_{n=1}^{N_{\rm real}}
    \left\langle
    \left|\delta G_{IJ,n}^{\kappa}(f_k)\right|^2
    \right\rangle \\
    &=
    \frac{1}{N_{\rm real}^2}
    \sum_{n=1}^{N_{\rm real}}
    P_{II}^{\kappa}(f_k)P_{JJ}^{\kappa}(f_k) \\
    &=
    \frac{P_{II}^{\kappa}(f_k)P_{JJ}^{\kappa}(f_k)}
    {N_{\rm real}} .
\end{aligned}
\end{equation}
Here the second equality follows from Eqs.~\eqref{tgm} and~\eqref{wkx}. Then we have
\begin{equation}
    \sigma\!\left(\widehat G_{IJ}^{\kappa}(f_k)\right)
    \equiv
    \left[
    \left\langle
    \left|\widehat G_{IJ}^{\kappa}(f_k)-P_{IJ}^{\kappa}(f_k)\right|^2
    \right\rangle
    \right]^{1/2}
    =
    \left[
    \frac{P_{II}^{\kappa}(f_k)P_{JJ}^{\kappa}(f_k)}
    {N_{\rm real}}
    \right]^{1/2}.
\end{equation}
For nonzero \(P_{IJ}^{\kappa}(f_k)\), dividing by \(|P_{IJ}^{\kappa}(f_k)|\) converts this absolute fluctuation into a relative complex residual. Introduce the matrix-element coherence
\begin{equation}\label{enn}
    \rho^{\kappa}_{IJ}(f_k)
    =
    \frac{|P^{\kappa}_{IJ}(f_k)|}
    {\sqrt{P^{\kappa}_{II}(f_k)P^{\kappa}_{JJ}(f_k)}} ,
    \qquad
    0\leq\rho^{\kappa}_{IJ}\leq1,
\end{equation}
so that
\begin{equation}
    \frac{
    \sigma\!\left(\widehat G_{IJ}^{\kappa}(f_k)\right)
    }{
    |P_{IJ}^{\kappa}(f_k)|
    }
    =
    \frac{1}
    {\sqrt{N_{\rm real}}\,\rho_{IJ}^{\kappa}(f_k)} .
\end{equation}

The diagnostic underlying Figs.~\ref{fig:diagonaluncertaintyscatter} and~\ref{fig:crosscoherencescatter} is the observed residual $r_{IJ}^{\kappa}$ defined in Eq.~\eqref{eq:crossspectrum_relative_residual}. The panels in Figs.~\ref{fig:diagonaluncertaintyscatter} and~\ref{fig:crosscoherencescatter} display $N_{\rm real}[r_{IJ}^{\kappa}]^2$; squaring the residual itself and taking the ensemble expectation value gives
\begin{equation}
\begin{aligned}
    \left\langle
    \left[r_{IJ}^{\kappa}(f_k)\right]^2
    \right\rangle
    &=
    \left\langle
    \frac{
    \left|\widehat G_{IJ}^{\kappa}(f_k)-P_{IJ}^{\kappa}(f_k)\right|^2
    }{
    \left|P_{IJ}^{\kappa}(f_k)\right|^2
    }
    \right\rangle 
    =
    \frac{
    P_{II}^{\kappa}(f_k)P_{JJ}^{\kappa}(f_k)
    }{
    N_{\rm real}\left|P_{IJ}^{\kappa}(f_k)\right|^2
    } 
    =
    \frac{1}
    {N_{\rm real}\,[\rho_{IJ}^{\kappa}(f_k)]^2},
\end{aligned}
\end{equation}
where the second and third equalities follow from Eqs.~\eqref{slx} and~\eqref{enn}, respectively.

\section{Full-Covariance SNR Expressions}\label{app:snr-derivation}

This appendix derives the SNR expressions quoted in Eq.~\eqref{eq:snrdefs}. Following Ref.~\cite{Chen:2024fto}, introduce an auxiliary amplitude \(A\) that rescales the covariance contribution of the target stochastic component,
\begin{equation}
\mathbf P^{\kappa}(f_k;A)
=
\mathbf P_{\rm base}^{\kappa}(f_k)
+
A\,\mathbf P_{\rm targ}^{\kappa}(f_k).
\end{equation}
Here \(\mathbf P_{\rm targ}\) is the target component and \(\mathbf P_{\rm base}\) is the reference covariance already present in the corresponding search. The physical unit-amplitude target corresponds to \(A=1\).

Under the weak-signal approximation \(A\,\mathbf P_{\rm targ}\ll \mathbf P_{\rm base}\), the target signal contributes only weakly to the total covariance, allowing the dependence on \(A\) to be approximated by an expansion around \(\mathbf P_{\rm base}\). Substituting this one-parameter model into the local FIM expression in Eq.~\eqref{eq:fim} gives
\begin{equation}\label{eq:amplitude_fim}
F_{AA}
=
\sum_{\kappa,k}
\operatorname{Tr}\!\left[
(\mathbf P_{\rm base}^{\kappa}(f_k))^{-1}
\mathbf P_{\rm targ}^{\kappa}(f_k)
(\mathbf P_{\rm base}^{\kappa}(f_k))^{-1}
\mathbf P_{\rm targ}^{\kappa}(f_k)
\right].
\end{equation}
In this local one-parameter construction, the amplitude variance is
\begin{equation}
\sigma_A^2 \simeq F_{AA}^{-1},
\end{equation}
and the SNR is
\begin{equation}
\mathrm{SNR}=\frac{\langle \widehat A\rangle}{\sigma_A}.
\end{equation}
For a normalized target template with \(\langle\widehat A\rangle=1\), this gives \(\mathrm{SNR}^2=F_{AA}\).

The absolute SNR is obtained by taking the baseline covariance in Eq.~\eqref{eq:amplitude_fim} to be \(\mathbf P_{\rm noise}\). The relative SNR is obtained by taking the baseline covariance to be the search-specific \(\mathbf P_{\rm base}\). Using the same one-sided positive-frequency bin convention as Eqs.~\eqref{eq:fim} and~\eqref{eq:snrdefs}, this gives
\begin{equation}
\begin{aligned}
\mathrm{SNR}_{\rm abs}^2
&=
\sum_{\kappa,k}
\operatorname{Tr}\!\left[
\left(\mathbf P_{\rm noise}^{\kappa}(f_k)\right)^{-1}
\mathbf P_{\rm targ}^{\kappa}(f_k)
\left(\mathbf P_{\rm noise}^{\kappa}(f_k)\right)^{-1}
\mathbf P_{\rm targ}^{\kappa}(f_k)
\right],
\\
\mathrm{SNR}_{\rm rel}^2
&=
\sum_{\kappa,k}
\operatorname{Tr}\!\left[
\left(\mathbf P_{\rm base}^{\kappa}(f_k)\right)^{-1}
\mathbf P_{\rm targ}^{\kappa}(f_k)
\left(\mathbf P_{\rm base}^{\kappa}(f_k)\right)^{-1}
\mathbf P_{\rm targ}^{\kappa}(f_k)
\right].
\end{aligned}
\end{equation}

In the equal-arm approximation, the \(XYZ\)-basis PSD matrices have the same real symmetric channel structure and are diagonalized by the fixed \(A,E,T\) rotation. Let \(\mathbf S\) denote this unitary transformation. For \(Q\in\{{\rm noise},{\rm base},{\rm targ}\}\),
\begin{equation}
    \begin{gathered}
    \mathbf S^\dagger \mathbf P_Q^\kappa(f_k)\mathbf S
    =
    \operatorname{diag}\!\left(
    P_{Q,A}^\kappa(f_k),
    P_{Q,E}^\kappa(f_k),
    P_{Q,T}^\kappa(f_k)
    \right),
    \\
    \mathbf S^\dagger \left(\mathbf P_Q^\kappa(f_k)\right)^{-1}\mathbf S
    =
    \left[\mathbf S^\dagger \mathbf P_Q^\kappa(f_k)\mathbf S\right]^{-1}.
    \end{gathered}
\end{equation}
Because the trace is invariant under cyclic unitary transformations, the matrix product inside each trace can be evaluated in this diagonal basis. Suppressing the common \((\kappa,f_k)\) arguments inside the trace, for example,
\begin{equation}
\begin{aligned}
\operatorname{Tr}\!\left[
(\mathbf P_{\rm base}^{\kappa})^{-1}
\mathbf P_{\rm targ}^{\kappa}
(\mathbf P_{\rm base}^{\kappa})^{-1}
\mathbf P_{\rm targ}^{\kappa}
\right]
&=
\operatorname{Tr}\!\left[
(\mathbf S^\dagger\mathbf P_{\rm base}^{\kappa}\mathbf S)^{-1}
(\mathbf S^\dagger\mathbf P_{\rm targ}^{\kappa}\mathbf S)
(\mathbf S^\dagger\mathbf P_{\rm base}^{\kappa}\mathbf S)^{-1}
(\mathbf S^\dagger\mathbf P_{\rm targ}^{\kappa}\mathbf S)
\right] \\
&=
\sum_{I=A,E,T}
\left[
\frac{P_{{\rm targ},I}^{\kappa}(f_k)}
{P_{{\rm base},I}^{\kappa}(f_k)}
\right]^2 .
\end{aligned}
\end{equation}
Thus the full-covariance \(XYZ\) expression reduces to the standard diagonal \(AET\)-channel form,
\begin{equation}
\begin{aligned}
\mathrm{SNR}_{\rm abs}^2
&=
\sum_{\kappa,k}\sum_{I=A,E,T}
\left[
\frac{P_{{\rm targ},I}^{\kappa}(f_k)}
{P_{{\rm noise},I}^{\kappa}(f_k)}
\right]^2,
\\
\mathrm{SNR}_{\rm rel}^2
&=
\sum_{\kappa,k}\sum_{I=A,E,T}
\left[
\frac{P_{{\rm targ},I}^{\kappa}(f_k)}
{P_{{\rm base},I}^{\kappa}(f_k)}
\right]^2 .
\end{aligned}
\end{equation}

\end{appendices}

\section*{Acknowledgments}
We would like to thank Gang Wang, Hansong Zhang and Renate Meyer for helpful discussions. H.-K. G. is supported by the startup fund provided by the University of Chinese Academy of Sciences and by the National Science Foundation of China (NSFC) under Grant No. 12547104 and No. 12475109. M. D. is supported by National Key Research and Development Program of China Grant No. 2021YFC2201903

\vspace{1.5\baselineskip}

\bibliographystyle{JHEP}
\bibliography{refer}

\end{document}